\documentclass[12pt]{article}
\usepackage{amsmath}
\usepackage{graphicx}
\usepackage{natbib}
\usepackage{url} % not crucial - just used below for the URL 

\newcommand{\blind}{0}

\DeclareMathOperator*{\argmax}{arg\,max}

\DeclareMathSymbol{\mh}{\mathord}{operators}{`\-}
\usepackage{amsmath}
\usepackage{dsfont}
\usepackage[colorlinks,citecolor=blue,urlcolor=blue,filecolor=blue,backref=page]{hyperref}

\usepackage{algorithm} 
\usepackage{algpseudocode}
\usepackage[graphicx]{realboxes} %rotate the sideways tables
\usepackage{caption}
\usepackage{comment}
\usepackage{subcaption}
\usepackage{booktabs}
\usepackage{amsfonts}
\usepackage{pdflscape}
\usepackage{tabularx}
\usepackage[Figuresright]{rotating}
\usepackage{graphicx}
\graphicspath{{./img/}}

\DeclareUnicodeCharacter{2212}{-} 
\DeclareUnicodeCharacter{2113}{+} 

\begin{document}

\def\spacingset#1{\renewcommand{\baselinestretch}%
{#1}\small\normalsize} \spacingset{1}

%%%%%%%%%%%%%%%%%%%%%%%%%%%%%%%%%%%%%%%%%%%%%%%%%%%%%%%%%%%%%%%%%%%%%%%%%%%%%%

% \title{\bf A Data Driven Bayesian Graphical Ridge Estimator}
% \author{
% \name{J. Smith \textsuperscript{a},
% M. Arashi \textsuperscript{b}\thanks{CONTACT M.~Arashi. Email: m\_arashi\_stat@yahoo.com} \hspace{.001cm} and A. Bekker \textsuperscript{a}} \vspace{.4cm}\\
% \textsuperscript{a} Department of Statistics, University of Pretoria, \\ Pretoria, 0002, South Africa \\
% \textsuperscript{b} Department of Statistics, Faculty of Mathematical Sciences, \\ Ferdowsi University of Mashhad, Mashhad, Iran
% }

% \maketitle
\if0\blind
{
  \title{\bf A naïve Bayesian graphical elastic net: driving advances in differential network analysis}
  \author{
    J. Smith \hspace{.2cm}\\
    Department of Statistics, University of Pretoria, \\ Pretoria, 0002, South Africa\\
    and \\
    A. Bekker \\
    Department of Statistics, University of Pretoria, \\ Pretoria, 0002, South Africa 
    and \\
    M. Arashi \thanks{CONTACT M.~Arashi. Email: arashi@um.ac.ir} \\
    Department of Statistics, Faculty of Mathematical Sciences, \\ Ferdowsi University of Mashhad, Mashhad, Iran \\
    }
  \maketitle
} \fi

\if1\blind
{
  \bigskip
  \bigskip
  \bigskip
  \begin{center}
    {\LARGE\bf Title}
\end{center}
  \medskip
} \fi

\begin{abstract}
\noindent Differential Networks (DNs), tools that encapsulate interactions within intricate systems, are brought under the Bayesian lens in this research. A novel naïve Bayesian adaptive graphical elastic net (BAE) prior is introduced to estimate the components of the DN. A heuristic structure determination mechanism and a block Gibbs sampler are derived. Performance is initially gauged on synthetic datasets encompassing various network topologies, aiming to assess and compare the flexibility to those of the Bayesian adaptive graphical lasso and ridge-type procedures. The naïve BAE estimator consistently ranks within the top two performers, highlighting its inherent adaptability. Finally, the BAE is applied to real-world datasets across diverse domains such as oncology, nephrology, and enology, underscoring its potential utility in comprehensive network analysis.
\end{abstract}

\noindent%
\vfill

\newpage
\spacingset{1.5} % DON'T change the spacing!
\section{Introduction}\label{sec:introduction}
Gaussian graphical models (GGMs) offer a network-based analytical framework for multivariate Gaussian data. In this context, each node in the undirected graph represents one of $p$ variables, and weighted edges reflect the strength of the partial correlation between variable pairs. An absence of an edge between nodes equates to zero partial correlation, implying conditional independence between two nodes, given all the others in the network. The goal when estimating a GGM is to evaluate the corresponding precision matrix $\mathbf{\Omega}=\mathbf{\Sigma}^{-1}$, the inverse of the covariance matrix, using the sampled multivariate Gaussian data \citep{lauritzen1996graphical}. \par

Several techniques exist in the literature for estimating GGMs, each offering different perspectives and handling the complexity of high-dimensional data in various ways. From a frequentist viewpoint, \citep{meinshausen2006high} introduces a technique based on sequential "nodewise" regressions, becoming a key player in sparse graph estimation for high-dimensional data. This is followed by the popular Glasso, an approach proposed by \citep{friedman_2008_sparse}, which leverages the use of a lasso $\ell_1$ penalty to estimate a sparse inverse covariance matrix. In addition to this, a myriad of techniques have been proposed, including but not limited to the 'CLIME' (constrained $\ell_1$ minimization estimator) method \citep{cai2011constrained}; a joint estimation procedure across multiple classes can be found in \citep{guo2011joint,danaher2014joint}. More recently, \citep{shutta2023spider} introduces the SpiderLearner, an ensemble method that constructs a consensus network from multiple estimated GGMs.

On the Bayesian side of methodologies, hierarchical priors are commonly employed for the dual tasks of precision matrix estimation and structure learning in GGMs, with the G-Wishart as the typical prior used for $\mathbf{\Omega}$ \citep{dawid_1993_hyper, roverato2002hyper, letac_2007_wishart}. The use of alternate priors has also gained popularity. For example, \citep{wang_2015_scaling} presents a structure learning approach for concentration and covariance graph models that improves scalability through the use of continuous spike and slab priors, achieving comparable accuracy to conventional methods but with substantially reduced computation time and simpler implementation. \citep{mohammadi_2015_bayesian} develops a novel graph-constrained Bayesian structure learning approach for sparse Gaussian graphical models, showcasing superior performance in edge recovery and precision matrix estimation. Additional Bayesian methods that employ search algorithms across the graph space for simultaneous graph structure and precision matrix estimation encompass those but not limited to the following \citep{cheng_2012_hierarchical, mohammadi2017bayesian, mohammadi2021accelerating, van2022g}. Moreover, \citep{williams2021bayesian} unfolds a holistic Bayesian framework for Gaussian graphical models, encompassing shrinkage priors-enhanced structure learning, predictability augmentation, and network comparisons. Distinct from the G-Wishart prior approach, other Bayesian methodologies prioritize posterior mode estimation using shrinkage priors that avoid assigning positive probability mass at zero for off-diagonal elements, thereby boosting computational scalability through the application of efficient algorithms such as block Gibbs samplers. Specifically, independent exponential and Laplace priors for diagonal and off-diagonal elements respectively, form the basis of the Bayesian graphical lasso described by \citep{wang_2012_efficient}. Similarly, in \citep{smith2022data}, a block Gibbs sampler is devised, making use of a ridge-type penalisation characteristic of the graphical lasso method. \par

Networks, serving as representations of interactions within complex systems, pervade numerous scientific fields, providing insights into the system's behavior. Differential network (DN) analysis, the process of comparing networks across time or different states, further enriches this understanding by illuminating how these interactions evolve or respond to alterations in system conditions \citep{ideker2012differential}. A rigorous statistical perspective is provided in \citep{ali_shojaie}. The task of estimating DN ($\mathbf{\Delta}=\mathbf{\Omega_2}-\mathbf{\Omega_1}$) increasingly gains attention, and methods can be broadly categorized into two main types. The first approach estimates the precision matrices simultaneously. For example, \citep{zhao2014direct} provides a method for direct $\mathbf{\Delta}$ estimation, bypassing the need for individual precision matrices, while \citep{yuan2017differential} and \citep{jiang2018direct} use an alternating direction method of multipliers algorithm for $\mathbf{\Delta}$ estimation from a joint $\ell_1$ penalised loss function, and \citep{tang2020fast} proposes a rapid iterative shrinkage-thresholding algorithm. The second methodology independently estimates the individual precision matrices, $\mathbf{\Omega_1}$ and $\mathbf{\Omega_2}$; the resultant DN is derived from the difference between these estimated precision matrices. For example, the GGM estimation techniques discussed above can be used directly to estimate $\mathbf{\Delta}$. For example, \citep{smith2022empowering} utilises the Bayesian adaptive graphical lasso to analyse the difference in structures between various phases of the COVID-19 pandemic in South Africa. \par
In this context, similarly to the aforementioned author, a DN estimation technique is developed. In particular, gaining inspiration from the naïve elastic net approach from \citep{zou2005regularization}, the Bayesian graphical ridge-type and Bayesian graphical lasso are used as a springboard in constructing a naïve Bayesian graphical elastic net for separately estimating the components of the DN. The goal of this formulation is to provide adaptive performance across various network topologies, eliminating the need to compromise on either sparsity or numerical accuracy in real world applications. A novel block Gibbs sampler is derived and implemented in the "baygel" R package. \par

The remainder of this paper is organized as follows. Section \ref{sec:preliminaries} introduces the notation and Bayesian formulation for estimating individual GGM components. Section \ref{sec:the_bayesian_graphical_elastic_net} defines the naïve Bayesian graphical elastic net prior, its adaptive variant, and presents a heuristic structure determination mechanism and an associated block Gibbs sampler. Synthetic data applications, utilized to evaluate the novel estimator, are outlined in Section \ref{sec:synthetic_examples}. Finally, Section \ref{sec:application_study} demonstrates the application of the Bayesian adaptive elastic net estimator on diverse oncology, nephrology, and enology datasets. Finally, the manuscript concludes with a brief discussion and review of performance, limitations, and areas of future work.

\section{Preliminaries}\label{sec:preliminaries}{
This section aims to establish the necessary groundwork for the exploration of DN estimation. This includes a concise overview of the key notations, definitions, and mathematical formulations used in the process of estimating individual precision matrices, namely undirected GGMs \citep{lauritzen1996graphical}. This foundational knowledge equips the reader to follow and engage with the processes and methodologies involved.\par
Let $\mathcal{G}=(\mathcal{V}, \mathcal{E})$ define an undirected graphical model where $\mathcal{V}=\{1,2,...,p\}$ is the set of nodes and $\mathcal{E} \subseteq \mathcal{V} \times \mathcal{V}$ the set of existing edges. Consider the observations $\mathbf{y}=(\mathbf{y}_1, \mathbf{y}_2, ... ,\mathbf{y}_{n})$ to be an independent and identically distributed sample from a zero centered Gaussian distribution with covariance matrix $\mathbf{\Sigma}$. The GGM with respect to the graph $\mathcal{G}$ can be defined as $\mathcal{W}_{\mathcal{G}} = \big\{\mathrm{N_p}\left(\mathbf{0},\mathbf{\Sigma}\right)\:|\: \mathbf{\Omega}=\mathbf{\Sigma}^{-1} \in \mathbb{M}^{+}\big\}$, where $M^+$ is the space of positive definite matrices \citep{mohammadi_2015_bayesian}. Recall that the object of the graphical lasso is to maximize the penalized log-likelihood

\begin{equation} \label{eq:graphical_lasso}
   \argmax_{\mathbf{\Omega}\in \mathbb{M}^{+}} 
    \bigg\{\log(\mathrm{det}\mathbf{\Omega})-\mathrm{trace}\left(\frac{\mathbf{S}}{n}\mathbf{\Omega}\right) - \rho\Arrowvert \mathbf{\Omega} \Arrowvert_1\bigg\},
\end{equation}
}

\noindent where, $\rho \geq 0$ is the shrinkage parameter \citep{friedman_2008_sparse}. A fully Bayesian treatment of \eqref{eq:graphical_lasso}, namely the Bayesian graphical lasso is provided by

\begin{equation} \label{eq:bayes_glasso_prior}
    p\left(\mathbf{\Omega}\:|\lambda\right)=C^{-1}\prod_{i<j}\bigg\{\mathrm{DE}(\omega_{ij}\:|\:\lambda)\bigg\}
    \prod_{i=1}^{p}\bigg\{\mathrm{EXP}(\omega_{ii}\:|\:\lambda)\bigg\}\mathds{1}_{\mathbf{\Omega}\in \mathbb{M}^{+}}.
\end{equation}

\noindent The Bayesian adaptive graphical lasso is a generalisation of \eqref{eq:bayes_glasso_prior} which aims to provide adaptive shrinkage for each off diagonal element of $\mathbf{\Omega}$  \citep{wang_2012_bayes}

\begin{equation}\label{eq:bayesian_adaptive_lasso_prior}
    \begin{aligned}
        p(\mathbf{\Omega} \:|\: \{\lambda_{ij}\}_{i\leq j})&= 
    C^{-1}_{\{\lambda_{ij}\}_{i\leq j}}
    \prod_{i<j}\bigg\{\mathrm{DE}(\omega_{ij}\:|\:\lambda_{ij})\bigg\}
    \prod_{i=1}^{p}\bigg\{\mathrm{EXP}(\omega_{ii}\:|\:\frac{\lambda_{ii}}{2})\bigg\}\mathds{1}_{\mathbf{\Omega}\in \mathbb{M}^{+}}\\
        p(\{\lambda_{ij}\}_{i<j}\:|\:\{\lambda_{ii}\}_{i=1}^p)& \propto
    C_{\{\lambda_{ij}\}_{i\leq j}}
    \prod_{i<j}\mathrm{GA}(r,s).
    \end{aligned}
\end{equation}

\noindent The set of priors make use of the product of a double exponential (DE) with form $p(y)=\lambda/2\exp(-\lambda|y|)$ for the off diagonal elements and an exponential (EXP) with form $p(y)=\lambda \exp(-\lambda y)\mathds{1}_{y>0}$ for the diagonal. Secondly, different shrinkage parameters $\lambda_{ij}$ are placed on each off-diagonal element $\omega_{ij}$.  Thirdly, $\{\lambda_{ii}\}_{i=1}^p$ for the diagonal elements are treated as hyperparameters. $\mathrm{GA}\left(r,s\right)$ represents a gamma density function with form $p(y)= s^r/\Gamma(r)y^{r-1}\mathrm{exp}\left(-sy\right)$ and the indicator function $\mathds{1}_{\mathbf{\Omega}\in \mathbb{M}^{+}}$ equals 1 if the condition is met and 0 otherwise. Finally, $C_{\{\lambda_{ij}\}_{i\leq j}}$ is the uncomputable normalization constant dependent on $\lambda_{ij}$, however, these constants will cancel out in the formulation of the posterior distribution - reducing the computational burden when updating $\lambda_{ij}$.\par

The authors in \citep{smith2022data} consider using an $\ell_2$ constraint in place of the $\ell_1$ within \eqref{eq:graphical_lasso}. This formulation aims to address accurate numerical estimation of non-sparse precision matrices over sparse representations and results in a graphical ridge-type problem where the objective is to maximise the log-likelihood

\begin{equation}  \label{eq:graphical_ridge}
   \argmax_{\mathbf{\Omega}\in \mathbb{M}^{+}} 
    \bigg\{
    \log(\mathrm{det}\mathbf{\Omega})-\mathrm{trace}\left(\frac{\mathbf{S}}{n}\mathbf{\Omega}\right) - \frac{\rho}{2}\Arrowvert \mathbf{\Omega} \Arrowvert_2^2
    \bigg\}.
\end{equation}

\noindent Using \eqref{eq:bayes_glasso_prior} as a departure point, the Bayesian analog of \eqref{eq:graphical_ridge} is given by the Bayesian graphical ridge-type prior,

\begin{equation} \label{eq:bayes_gridge_prior1}
    p\left(\mathbf{\Omega}\:|\mu=0,\sigma\right)=C^{-1}\prod_{i<j}\bigg\{\mathrm{N}(\omega_{ij}\:|\:\mu\:,\sigma)\bigg\}
    \prod_{i=1}^{p}\bigg\{\mathrm{TN}(\omega_{ii}\:|\:\mu\:,\sigma)\bigg\}\mathds{1}_{\mathbf{\Omega}\in \mathbb{M}^{+}}.
\end{equation}

\noindent Similar to \eqref{eq:bayesian_adaptive_lasso_prior}, the adaptive variant is given by

\begin{equation}\label{eq:bayesian_adaptive_ridge_prior}
    \begin{aligned}
        p\left(\mathbf{\Omega}\:|\mu=0,\{\tau_{ij}\}_{i<j}\right)&=C^{-1}_{\{\tau_{ij}\}_{i<j}}\prod_{i<j}\bigg\{\mathrm{N}(\omega_{ij}\:|\:\mu,\tau_{ij})\bigg\}
    \prod_{i=1}^{p}\bigg\{\mathrm{TN}(\omega_{ii}\:|\:\mu,\tau_{ii})\bigg\}\mathds{1}_{\mathbf{\Omega}\in \mathbb{M}^{+}}\\
        p\left(\{\tau_{ij}\}_{i<j}\:|\{\tau_{ii}\}^{p}_{i=1} \right)& \propto C_{\{\tau_{ij}\}_{i<j}}\prod_{i<j}\mathrm{IGA}(a,b).
    \end{aligned}
\end{equation}

\noindent Here, $\mathrm{N}(y\:|\sigma)$ represents a Gaussian density function and $\mathrm{TN}(y\:|\sigma)$ represents a univariate left truncated at zero Gaussian density function with form  $p(y)=\sqrt{2}(\sigma\sqrt{\pi})^{-1}\exp(-0.5[y/\sigma]^2)1_{y>0}$. Next, $\tau = \sigma^{2}$ and $\mathrm{IGA}(a,b)$ represents an inverse gamma distribution with a shape parameter $a$ and scale parameter $b$ and form $p(y)=b^a/\Gamma(a)y^{a+1}\mathrm{exp}(-b/y)$. Finally, $\{\tau_{ii}\}_{i=1}^p$ for the diagonal elements are also treated as hyperparameters and the same technique as before is used for the evaluation of the intractable normalising constant $C_{\{\tau_{ij}\}_{i<j}}$.

\section{The naïve Bayesian graphical elastic net}\label{sec:the_bayesian_graphical_elastic_net}
\subsection{The elastic net prior}\label{subsec:the_elastic_net_prior}
Considering the extensive array of possible network topologies, each with its unique degree of sparsity and magnitude, selecting the most suitable estimation technique can be a complex task. The heterogeneity inherent in these topologies necessitates an adaptable estimation approach, capable of assimilating strengths from various techniques \citep{shutta2023spider}. In the frequentist domain, the elastic net penalty enjoys success in estimating the sparsity pattern and edge weights in GGMs \citep{bernardini2022new}. Taking \eqref{eq:graphical_lasso} and \eqref{eq:graphical_ridge} as our starting point, we can embark on the journey of precision matrix estimation by combining the regularisation techniques from \eqref{eq:bayesian_adaptive_lasso_prior} and \eqref{eq:bayesian_adaptive_ridge_prior}. In particular, following the approach of \citep{zou2005regularization}, a naïve graphical elastic net can be formulated by combining the $\ell_1$ and $\ell_2$ constraints. To this end, the hierarchical form of the naïve Bayesian elastic net is given by

\begin{equation} \label{eq:bayes_graphical_BAE_prior}
    \begin{aligned}    
            p\left(\mathbf{y}_i\:|\mathbf{\Omega} \right)&=N_p\left(\mathbf{y}_i\:|\mathbf{0},\:\mathbf{\Omega}^{-1}\right) \:\:\: (i=1,\cdots,n)\\
            p\left(\mathbf{\Omega}|\mu=0,\sigma,\lambda \right)=&C_{*}^{-1}
            \prod_{i<j}\bigg\{\mathrm{DE}(\omega_{ij}|\lambda)\bigg\}
            \prod_{i=1}^{p}\bigg\{\mathrm{EXP}(\omega_{ii}|\lambda)\bigg\} \\
        &\times \prod_{i<j}\bigg\{\mathrm{N}(\omega_{ij}|\mu,\sigma)\bigg\}
    \prod_{i=1}^{p}\bigg\{\mathrm{TN}(\omega_{ii}|\mu,\sigma)\bigg\}\mathds{1}_{\mathbf{\Omega}\in \mathbb{M}^{+}}.
    \end{aligned}
\end{equation}

\noindent The normalising constant $C_{*}^{-1}$ in this form is not independent of $\sigma$ nor $\lambda$, however, techniques similar to those used in \eqref{eq:bayesian_adaptive_lasso_prior} and \eqref{eq:bayesian_adaptive_ridge_prior} can be useful here. The posterior is given by 

\begin{equation} \label{eq:bayes_BAE_posterior}
    \begin{aligned} 
        p(\mathbf{\Omega}|\mathbf{Y},\mu=0,\sigma,\lambda)
        &\propto
        \mathrm{det}\mathbf{\Omega}^{\frac{n}{2}}
        \exp\{-\mathrm{trace}(\frac{1}{2}\mathbf{S}\mathbf{\Omega})\} \\
        &\times
        \prod_{i<j}\bigg\{\frac{\lambda}{2}\exp(-\lambda|\omega_{ij}|)\bigg\}
        \prod_{i=1}^{p}\bigg\{\frac{\lambda}{2}\exp(-\frac{\lambda}{2}\omega_{ii})\bigg\}\\
        &\times
        \prod_{i<j}\bigg\{\frac{1}{\sigma\sqrt{2\pi}}\exp\left(-\frac{\omega_{ij}^2}{2\sigma^2}\right)\bigg\}
        \prod_{i=1}^{p}\bigg\{\frac{\sqrt{2}}{\sigma\sqrt{\pi}}\exp\left(-\frac{\omega_{ii}^2}{2\sigma^2}\right)\bigg\}\mathds{1}_{\mathbf{\Omega}\in \mathbb{M}^{+}}.
    \end{aligned}
\end{equation}

\noindent The log of the posterior is given by

\begin{equation}\label{eq:bayes_glasso_log_posterior}
        p_{\ell}(\mathbf{\Omega}|\mathbf{Y},\mu=0,\sigma,\lambda)
        \propto
        \log\left(\mathrm{det}\mathbf{\Omega}\right)-
        \mathrm{trace}\left(\frac{\mathbf{S}}{n}\mathbf{\Omega}\right)-
        \sum_{i<j}\frac{\lambda}{n}|\omega_{ij}|-
        \sum_{i=1}^{p}\frac{\lambda}{n}\omega_{ii}-
        \sum_{i<j}\frac{n}{\sigma^2}\omega_{ij}^2-
        \sum_{i=1}^{p}\frac{n}{\sigma^2}\omega_{ii}^2
\end{equation}

\noindent The value of $\mathbf{\Omega}$ which maximises \eqref{eq:bayes_glasso_log_posterior} is given by 

\begin{equation} \label{eq:graphical_elastic_net}
   \argmax_{\mathbf{\Omega}\in \mathbb{M}^{+}} 
    \log(\mathrm{det}\mathbf{\Omega})-\mathrm{trace}(\frac{\mathbf{S}}{n}\mathbf{\Omega}) - \rho_1\Arrowvert \mathbf{\Omega} \Arrowvert_1 - \rho_2\Arrowvert \mathbf{\Omega} \Arrowvert_2.
\end{equation}

\noindent where $\rho_1=\lambda /n$ and $\rho_2=n/ \sigma^2$. Finally, following in the footsteps of \citep{wang_2012_efficient} and \citep{park2008bayesian}, the double exponential distribution in \eqref{eq:bayes_graphical_BAE_prior} can be represented as a scale mixture of Gaussians leading to the hierarchical representation

\begin{equation}\label{eq:hierarchical_BAE_prior}
    \begin{aligned}
    p(\mathbf{\Omega}\:|\boldsymbol{\phi},\mu=0,\sigma,\lambda)&=C_{\boldsymbol{\phi}^*}^{-1}
    \prod_{i<j}\bigg\{\frac{1}{\sqrt{2\pi\phi_{ij}}}\exp(-\frac{\omega_{ij}^2}{2\phi_{ij}})\bigg\}
    \prod_{i=1}^{p}\bigg\{\frac{\lambda}{2}\exp(-\frac{\lambda}{2}\omega_{ii})\bigg\}\\
    &\times
    \prod_{i<j}\bigg\{\frac{1}{\sigma\sqrt{2\pi}}\exp\left(-\frac{\omega_{ij}^2}{2\sigma^2}\right)\bigg\}
    \prod_{i=1}^{p}\bigg\{\frac{\sqrt{2}}{\sigma\sqrt{\pi}}\exp\left(-\frac{\omega_{ii}^2}{2\sigma^2}\right)\bigg\}\mathds{1}_{\mathbf{\Omega}\in \mathbb{M}^{+}}\\
    p(\boldsymbol{\phi}|\mu=0,\sigma,\lambda) &\propto 
    C_{\boldsymbol{\phi}^*}\prod_{i<j}\frac{\lambda^2}{2}\exp(-\frac{\lambda^2}{2}\phi_{ij}).
    \end{aligned}
\end{equation}

\noindent Here, the intractable normalising constant, $C_{\boldsymbol{\phi}^*}$, depends on $\boldsymbol{\phi}$, $\lambda$ and $\sigma$.  

\begin{figure} [H]

\begin{minipage}{.333\linewidth}
\centering
\subfloat[]{\label{fig:marginal_density_omega_11}\includegraphics[scale=.28]{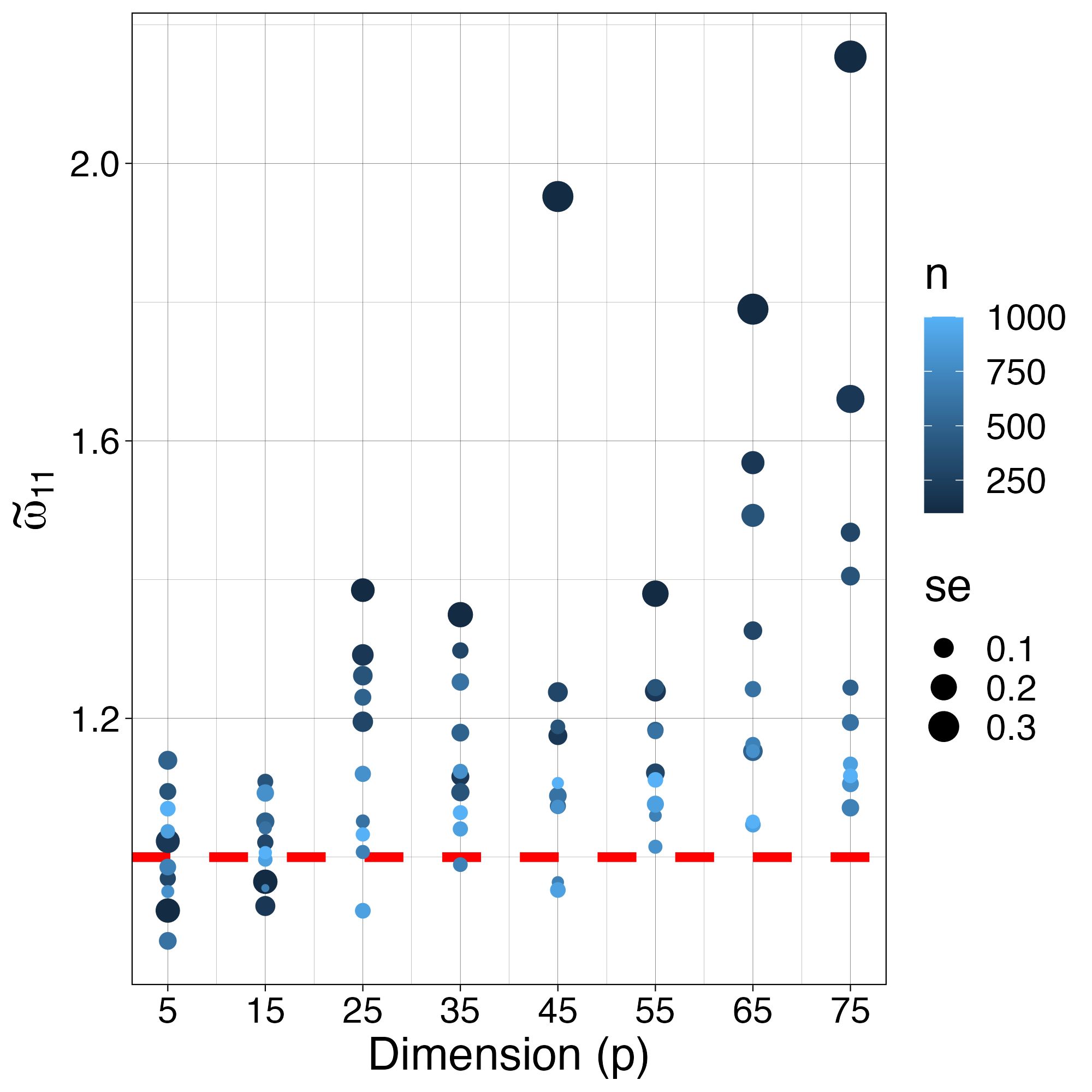}}
\end{minipage}%
\begin{minipage}{.333\linewidth}
\centering
\subfloat[]{\label{fig:tmarginal_density_omega_12}\includegraphics[scale=.28]{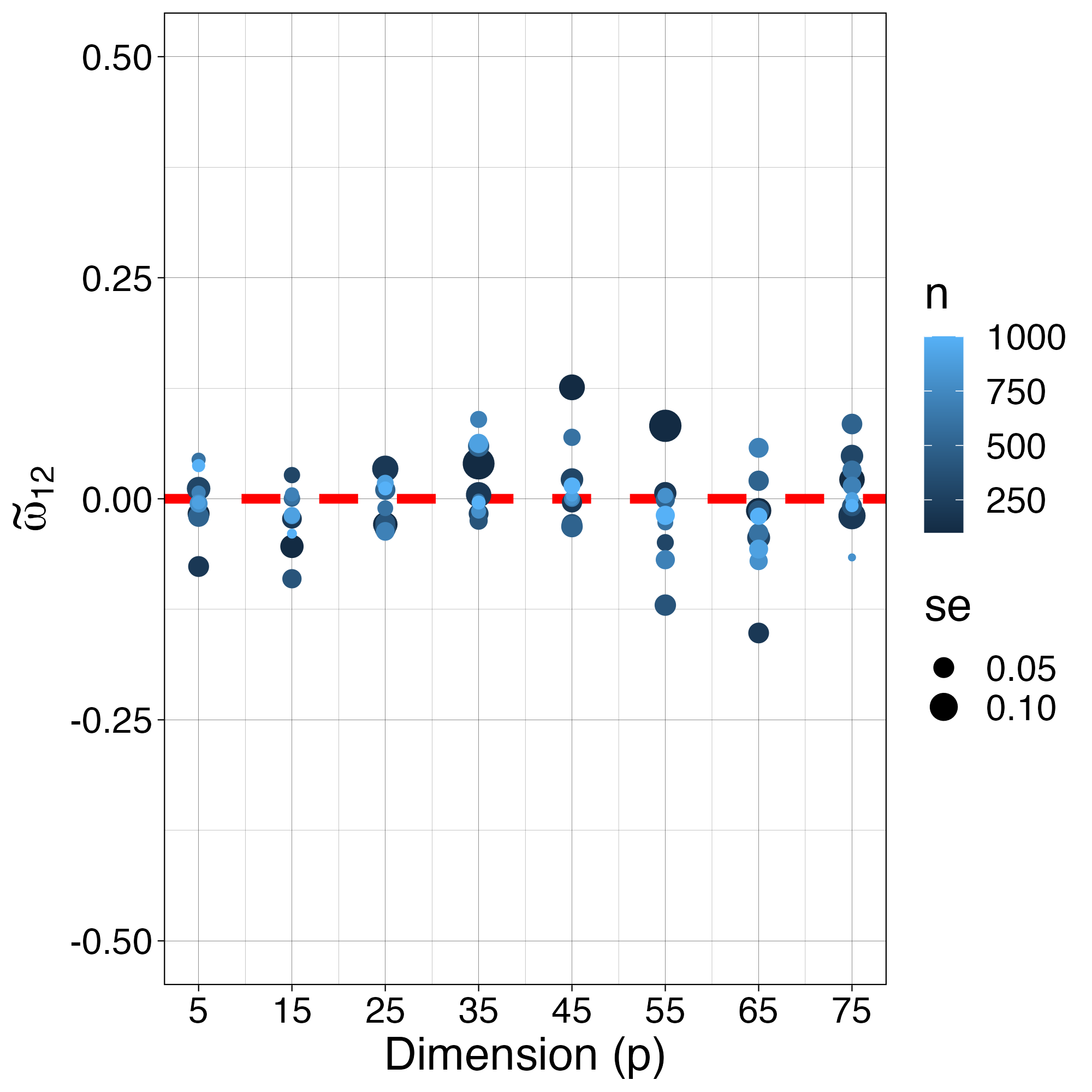}}
\end{minipage}
\begin{minipage}{.333\linewidth}
\centering
\subfloat[]{\label{fig:marginal_density_rho_12}\includegraphics[scale=.28]{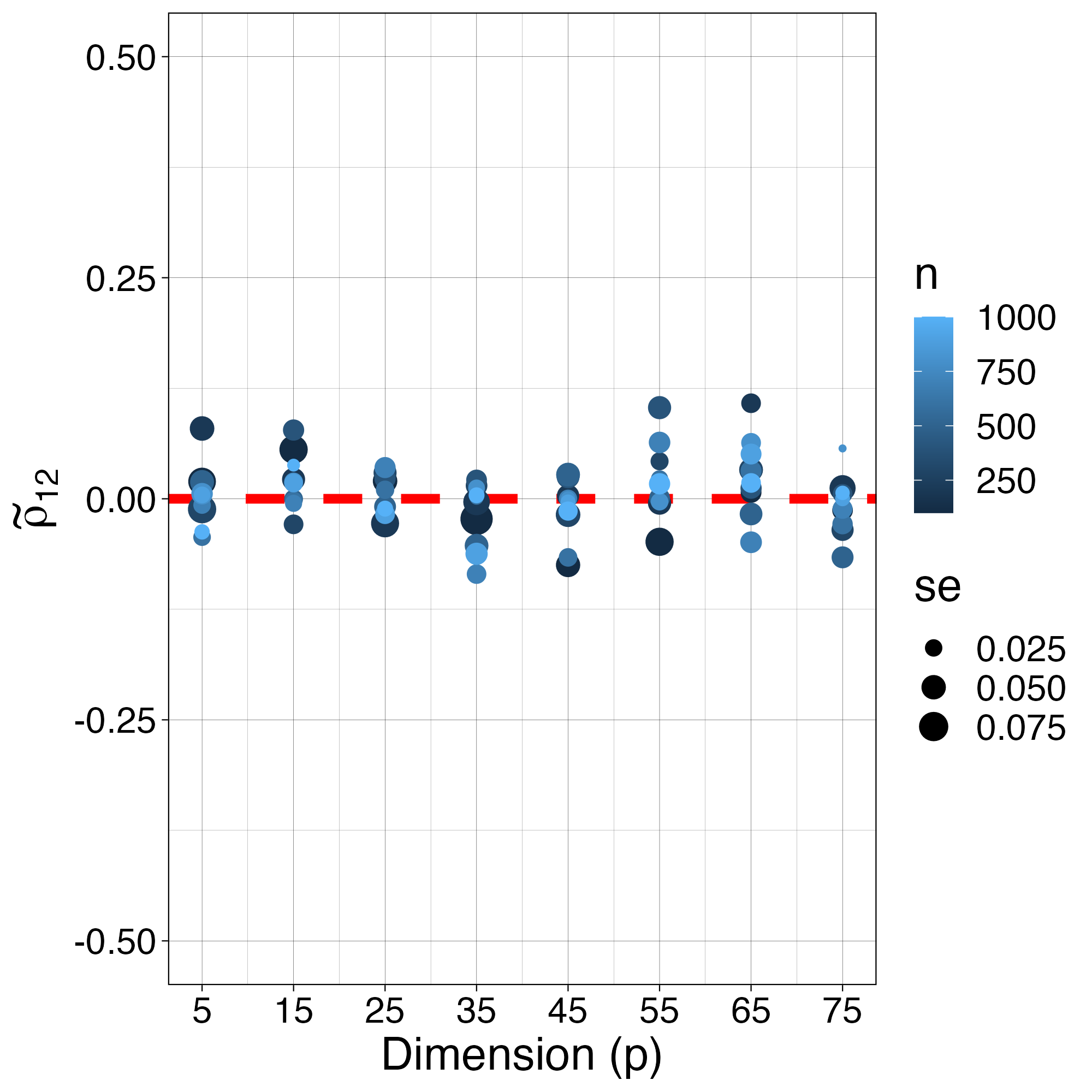}}
\end{minipage}

\caption{The distribution of the median for diagonal (a), off-diagonal (b), and partial correlation (c) elements of a diagonal $\mathbf{\Omega}$ matrix, with an increasing sample size ($n$) and $p$ ranges from 5 to 75. The magnitude of the dots corresponds to the standard error (se) of the elements, while the red line denotes the true value.}
\label{fig:marginal_densities}
\end{figure}

Figures (\ref{fig:marginal_density_omega_11}) - (\ref{fig:marginal_density_rho_12}) present the median distributions for specific elements of $\mathbf{\Omega}$, namely, a diagonal element, an off-diagonal element, and the associated partial correlation of the off-diagonal element. These distributions are being explored under conditions where $p$ varies between 5 and 75, and sample size $n$ ranges from 100 to 1000. The outcomes originate from synthetic samples created by the MCMC sampling mechanism outlined in Section \ref{subsec:block_gibbs_sampler_for_bayesian_elastic_net}. Notably, larger standard errors predominantly occur with smaller sample sizes and primarily for diagonal elements. There is a discernible upward trend in the distribution of the median of the diagonal elements as $p$ increases. In contrast, the median distributions for off-diagonal and partial correlation elements consistently cluster closely around zero across all $p$ values, a desirable trait inherited from the Bayesian graphical lasso. \par

\subsection{Structure learning}\label{subsec:bayesian_analysis_of_graph_structures}
When comprehensive Bayesian posterior analysis is required for the determination of network structure, it is essential that positive prior mass is assigned to events $\{\omega_{ij} = 0\}$. A key component in this methodology of Bayesian structure learning in GGMs is the prior distribution on $\mathbf{\Omega}$ given the graph $G$ constraints. The G-Wishart prior, \citep{roverato2002hyper}, is a popular choice for the conjugate prior of $\mathbf{\Omega}$, however, the stringent form and computational intensity of this approach could present significant challenges \citep{dobra_2011_bayesian, cheng_2012_hierarchical}. Contemporary Bayesian approaches employ variants of search algorithms within the graph space, capable of concurrently estimating both the graph structure and the precision matrix \citep{dobra2011copula,cheng_2012_hierarchical, wang_2012_efficient, lenkoski2013direct, mohammadi_2015_bayesian,mohammadi2017bayesian, mohammadi2021accelerating}. \par
In the frequentest domain, the traditional graphical lasso technique is capable of generating zeros for off-diagonal elements $(\omega_{ij})$ in the maximizer of the target function \eqref{eq:graphical_lasso} - providing a method for determining sparse network structures. Bayesian treatments of the graphical lasso model typically assign zero probability to the occurrence of ${\omega_{ij} = 0}$, thus resulting in zero posterior probability for this event \citep{wang_2012_bayes, smith2022empowering, smith2022data}. These techniques require heuristic approaches for graphical structure determination. For example, rooted in the thresholding approach described by \citep{carvalho2010horseshoe}, where \citep{wang_2012_bayes} claim ${\omega_{ij}=0}$ in the event

\begin{equation} \label{threshold_rule_wang}
   \frac{\Tilde{\rho}_{ij}}{E_g(\rho_{ij}\:|\:\mathbf{Y})} > 0.5.
\end{equation}

\noindent Here, $\Tilde{\rho}_{ij}$ is the expected value of the posterior sample partial correlation associated with the graphical lasso priors in \eqref{eq:bayes_glasso_prior} and $g$ is given by a standard conjugate Wishart $\mathrm{W}(3,\mathbf{I}_p)$ with parameter values selected based on evidence provided by \citep{jones2005experiments}. Similarly, \citep{smith2022empowering} suggests $\omega_{ij}\neq 0$ for $i \neq j$ if

\begin{equation} \label{threshold_rule_smith}
    |E_h(\rho_{ij}\:|\:\mathbf{Y})|>\eta,
\end{equation}\par

\noindent where $\eta$ minimises the absolute sparsity
error for each graph structure considered therein and $h$ the standard conjugate Wishart $\mathrm{W}(3,\epsilon\mathbf{I}_p)$ and $\epsilon = 0.001$. This approach necessitates a priori knowledge of the expected structure within applied modelling scenarios - this may not always be possible.\par

The advocated methodology here initiates with the latter approach, implementing a thresholding strategy underpinned by a data-driven endeavor. This aims to harmonize the numerical accuracy of precision matrix estimates while preserving sufficient adaptability in assembling network structures. To this end, the approach is relatively straight forward and suggests $\{\omega_{ij}\neq 0\}$ if

\begin{equation} \label{threshold_rule_smith_2023}
    |\Tilde{\rho}_{ij}|>\psi.
\end{equation}\par

\begin{figure}[H]
  \centering
   %\vspace{-10mm}
  \begin{tabular}[c]{ccc}
  \centering
  %\vspace{-10mm}
    \begin{subfigure}[c]{0.3\textwidth}
      \includegraphics[width=\textwidth]{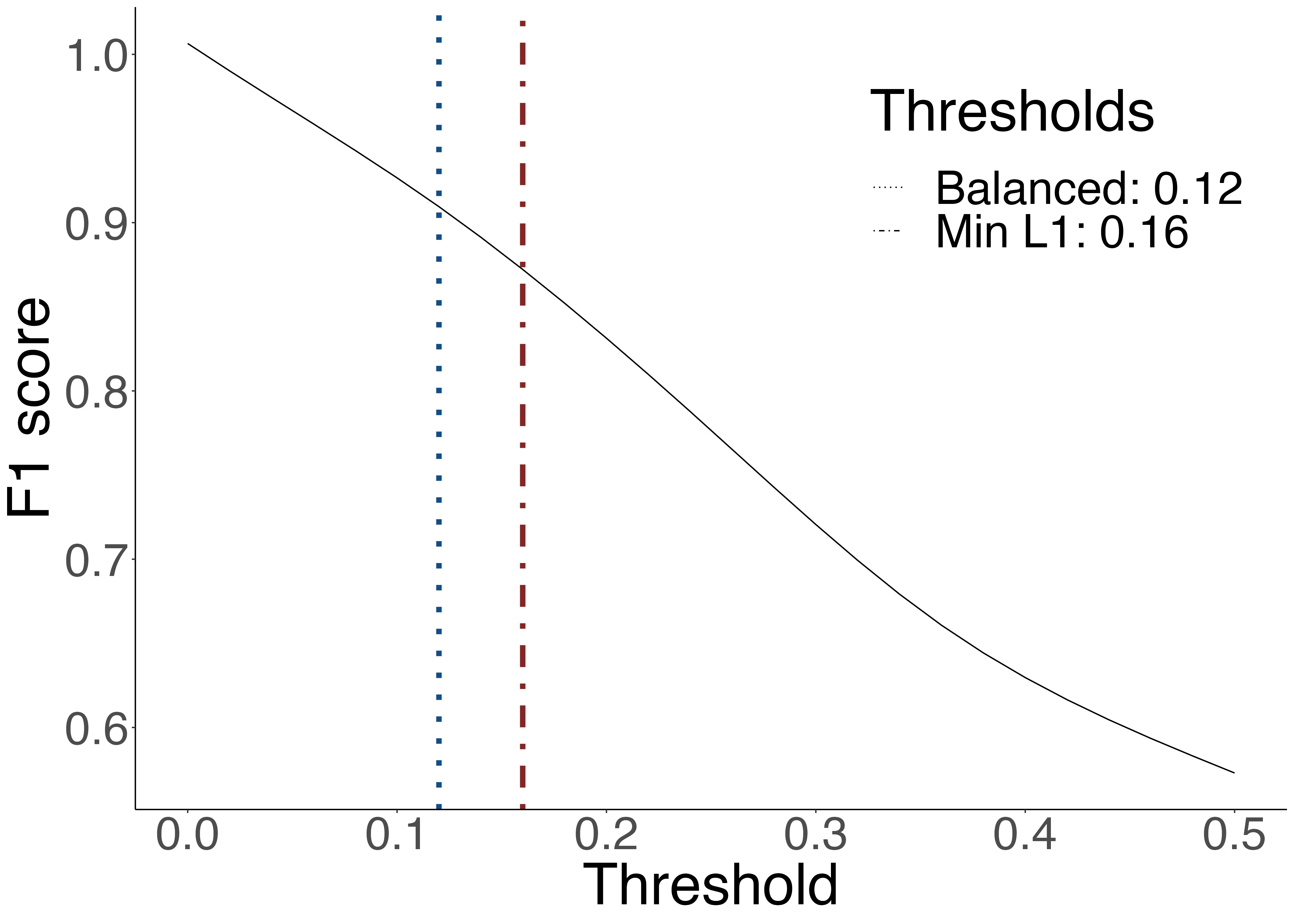}
         \caption{Model: M1.}
         \label{fig:thresh_study_p10_mod_1}
    \end{subfigure}&
    %\hspace{2mm  }
    \begin{subfigure}[c]{0.3\textwidth}
      \includegraphics[width=\textwidth]{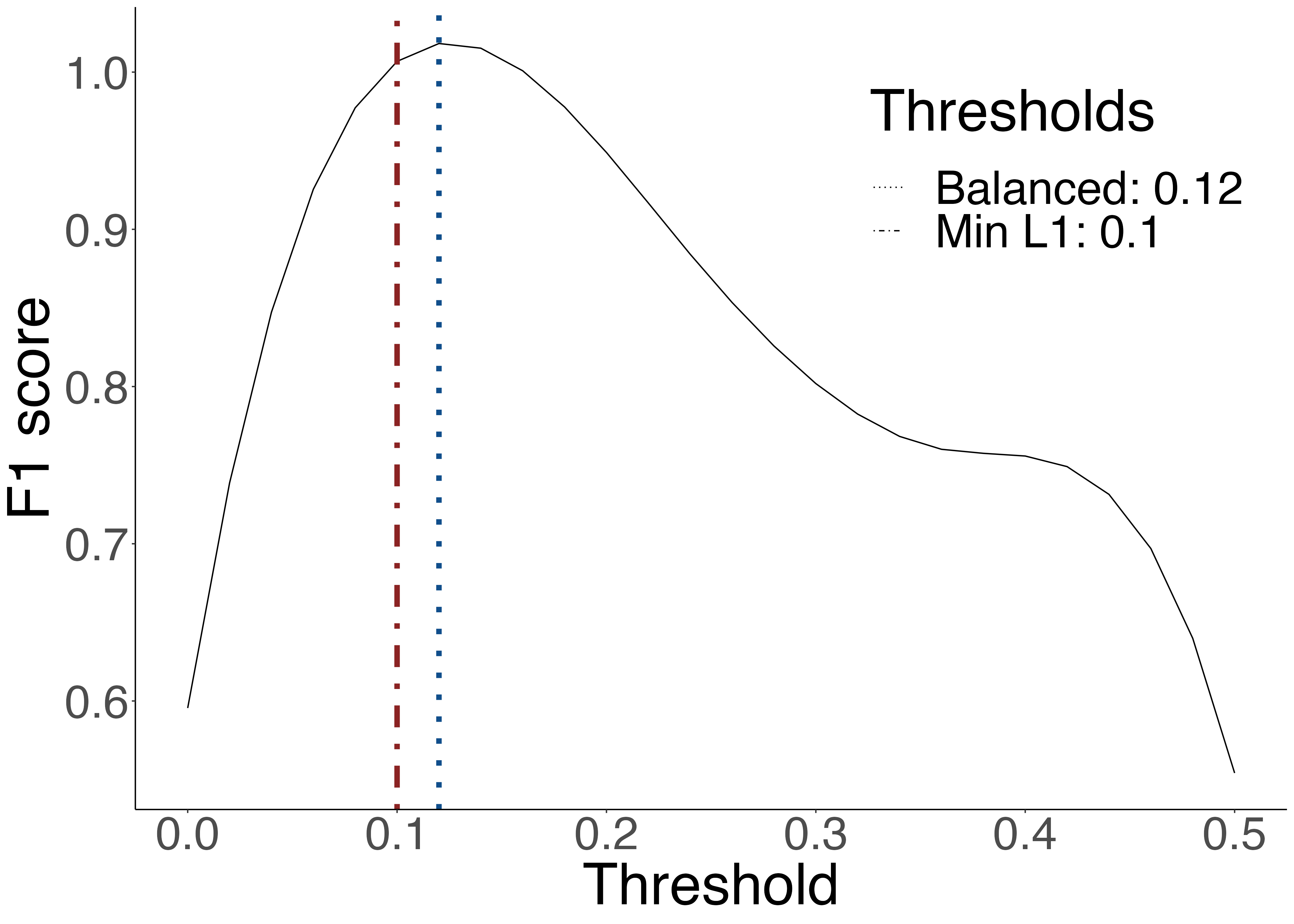}
         \caption{Model: M2.}
         \label{fig:thresh_study_p10_mod_2}
    \end{subfigure}&
    %\hspace{2mm  }
    \begin{subfigure}[c]{0.3\textwidth}
      \includegraphics[width=\textwidth]{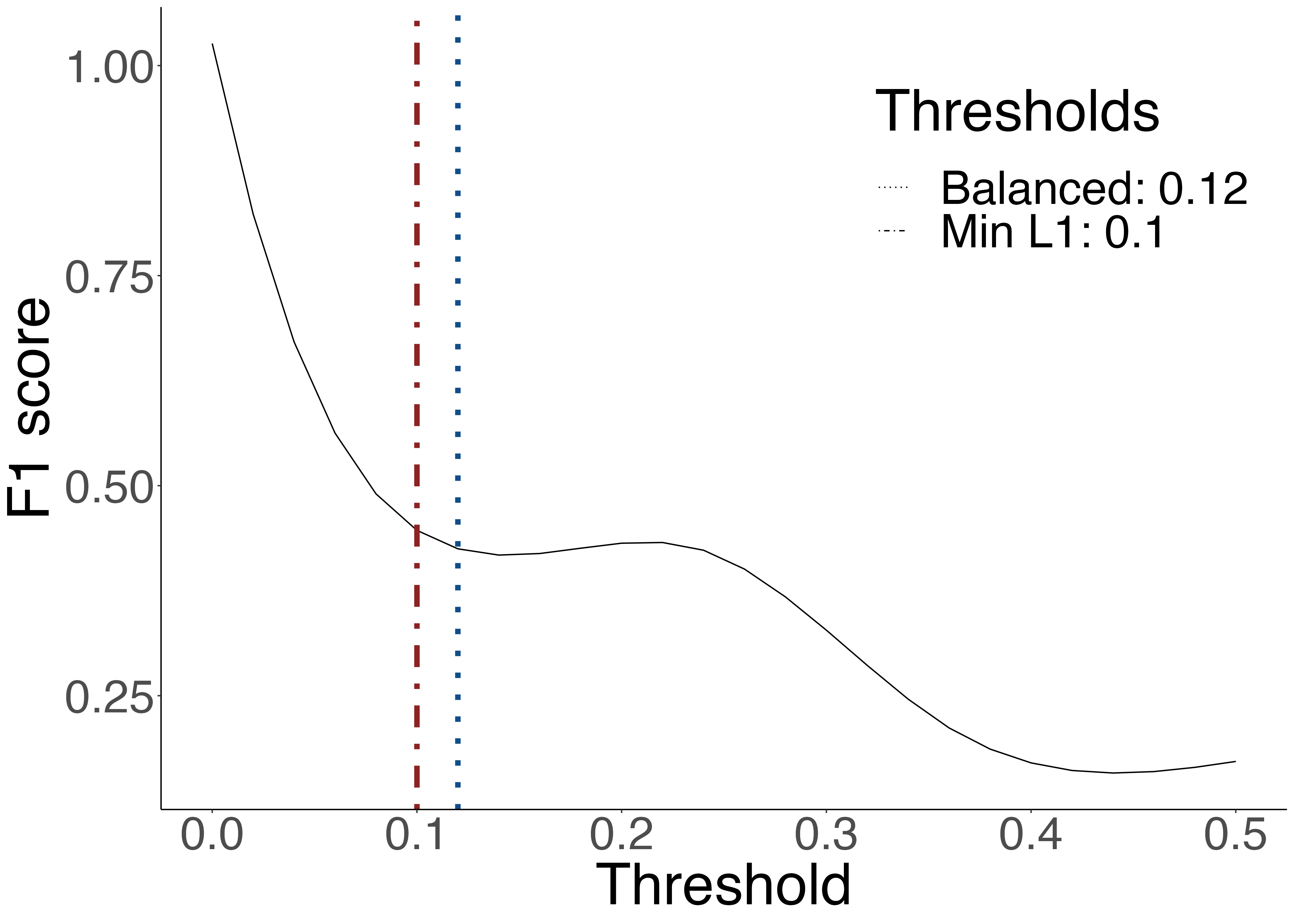}
         \caption{Model: M3.}
         \label{fig:thresh_study_p10_mod_3}
    \end{subfigure}\\
   % \vspace{-10mm}
    \begin{subfigure}[c]{0.3\textwidth}
      \includegraphics[width=\textwidth]{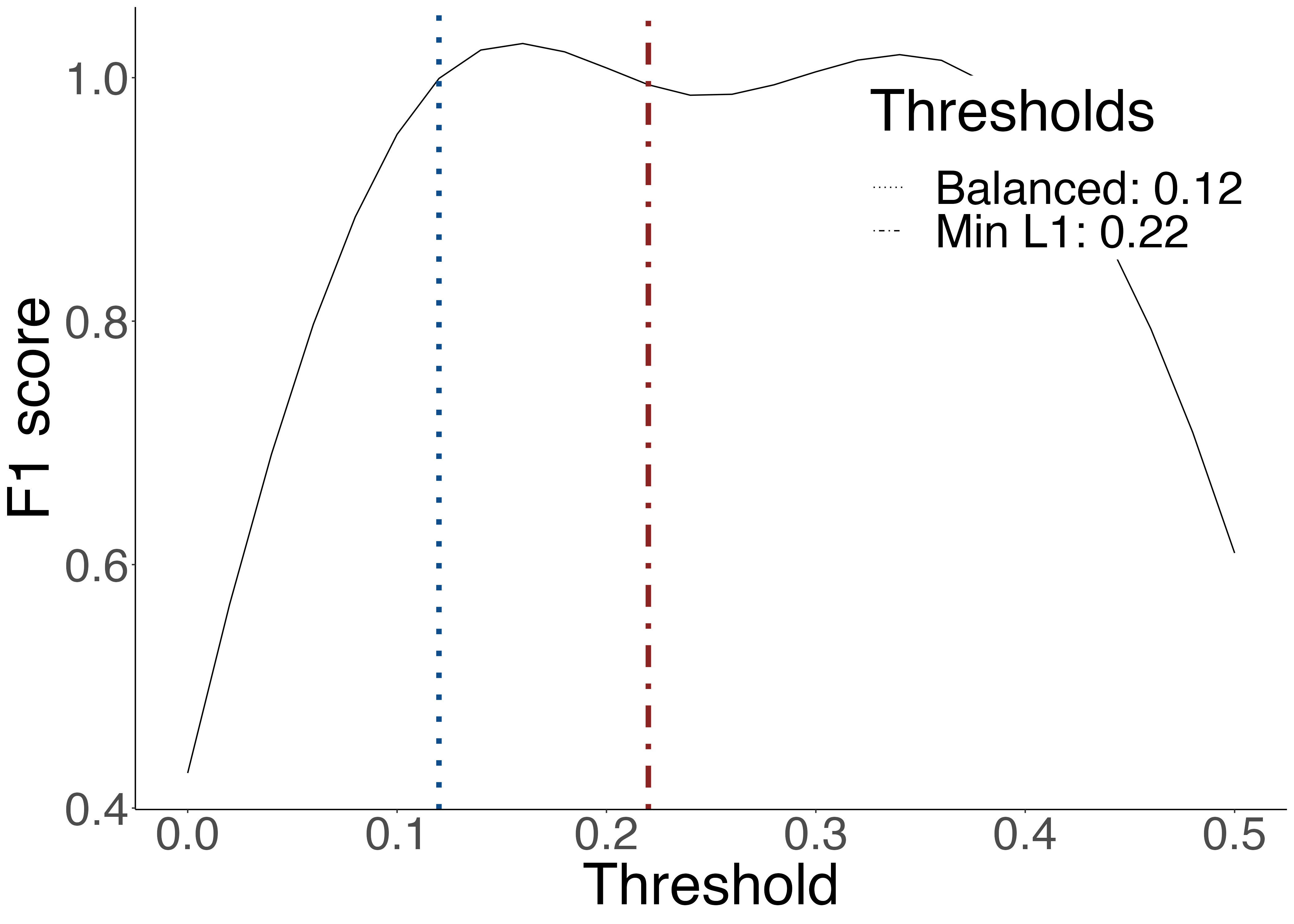}
        \caption{Model: M4.}
         \label{fig:thresh_study_p10_mod_4}
    \end{subfigure}&
    %\hspace{2mm  }
    \begin{subfigure}[c]{0.3\textwidth}
      \includegraphics[width=\textwidth]{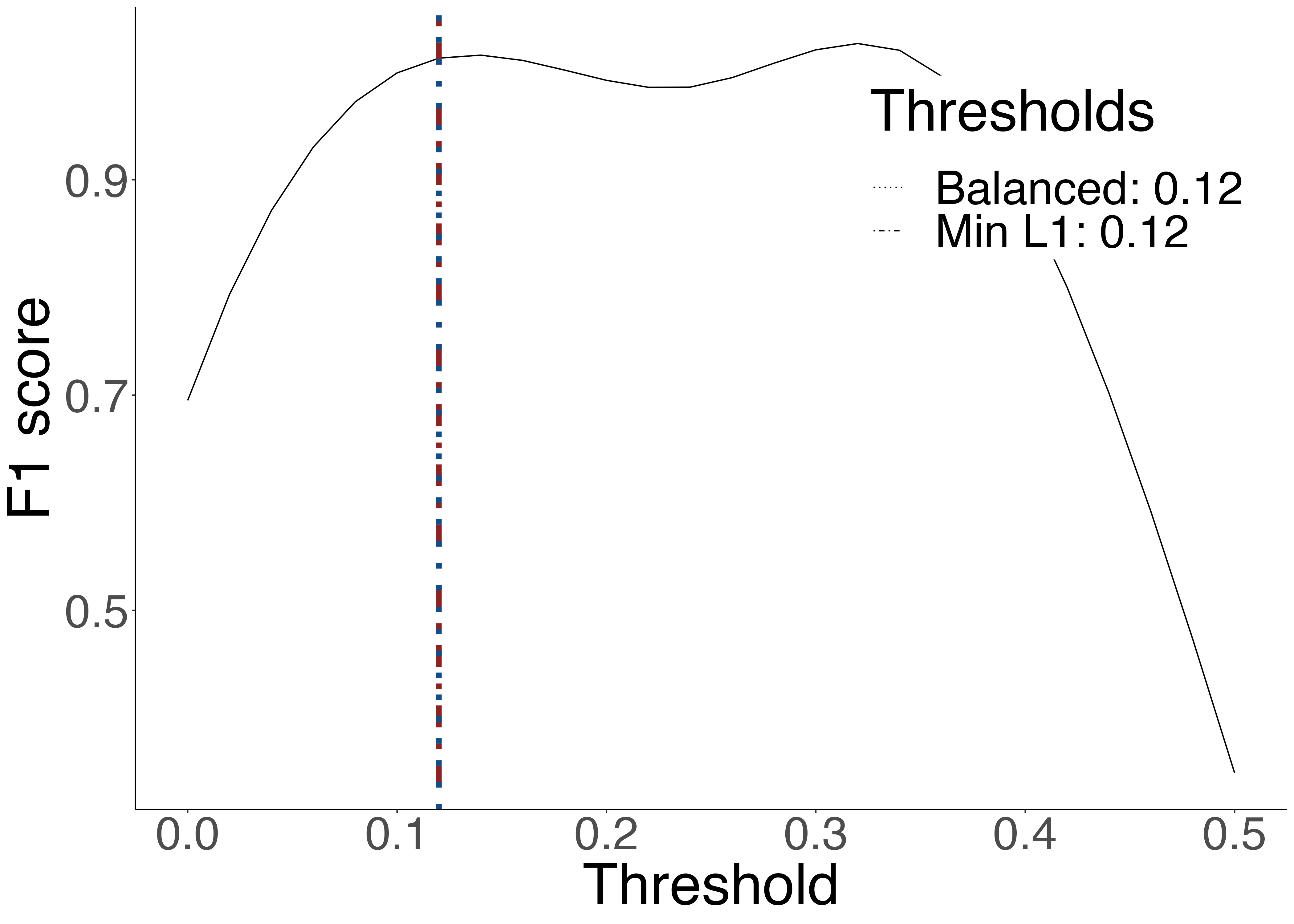}
        \caption{Model: M5.}
         \label{fig:thresh_study_p10_mod_5}
    \end{subfigure}&
    %\hspace{5mm  }
    \begin{subfigure}[c]{0.3\textwidth}
      \includegraphics[width=\textwidth]{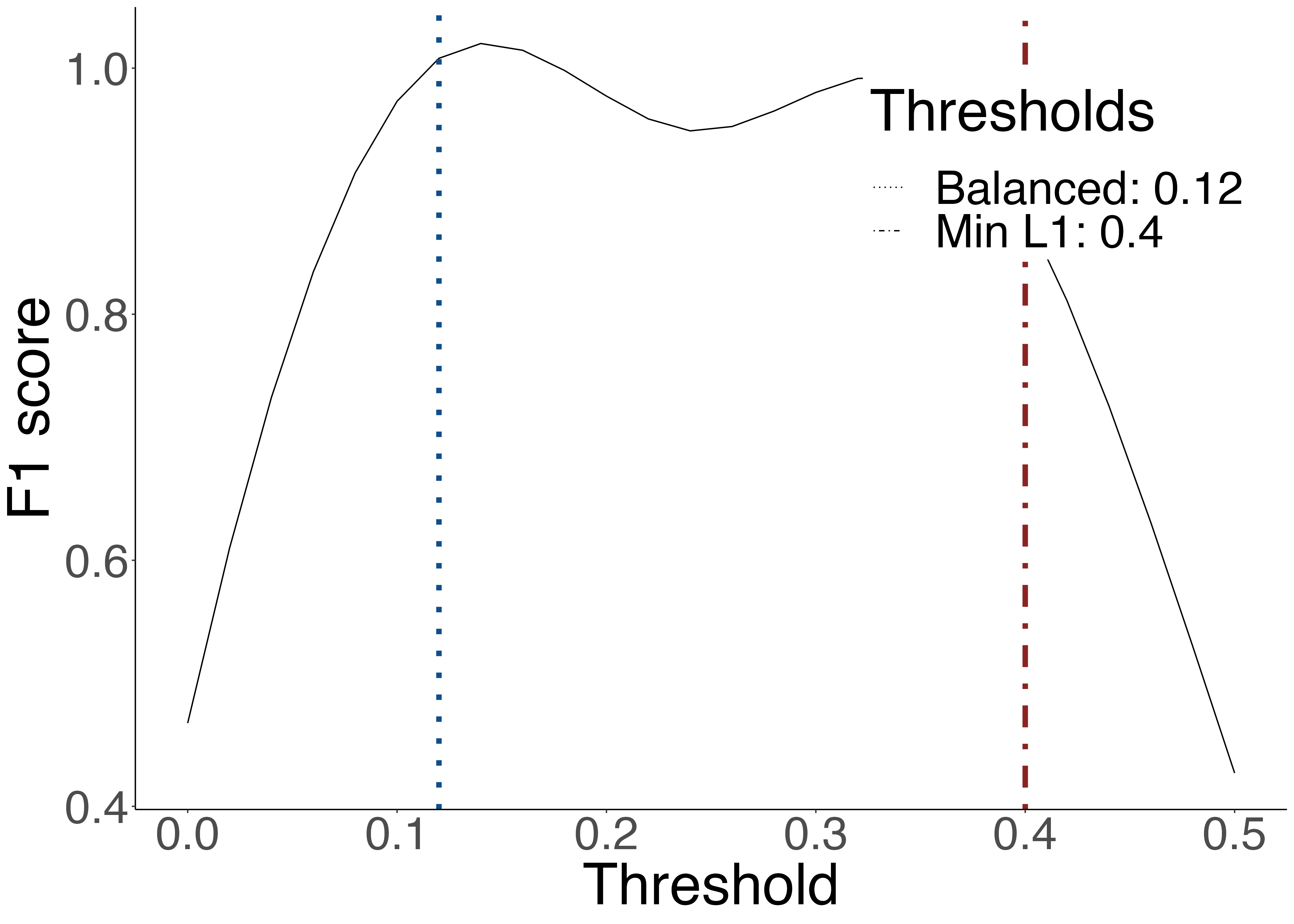}
        \caption{Model: M6.}
         \label{fig:thresh_study_p10_mod_6}
    \end{subfigure}
   \\
  \end{tabular} 
  %\hspace*{\fill}
    \caption{Assessment of F1 scores and L1 loss-minimizing thresholds across diverse network topologies for dimension $p=10$. Each subplot (M1 - M6) corresponds to a distinct model. Curves represent the trajectory of F1 scores against incrementing thresholds. The 'dot-dash' line indicates the threshold corresponding to the minimum L1 loss, while the 'dotted' line demonstrates the balanced threshold, $\psi = \pi_1\Tilde{\psi}_{f1} + \pi_2\Tilde{\psi}_{l1}$, which optimizes a balance between maximizing the F1 score and minimizing the L1 loss.} 
    \label{fig:elastic_net_threshold_study}
\end{figure}

The determination of $\psi$ entails an evaluation of thresholds $\psi_g$ that concurrently enhance the F1 score while diminishing the L1 score across a spectrum of distinct network topologies including but not limited to those described in Tables \ref{tab:syn_perf_functions} and \ref{tab:syn_studies_models}. These topologies exhibit unique characteristics and a range of edge densities. Finally, $\psi$ is  calculated as the equally weighted average of two median values - one representing thresholds that maximize F1 scores $\Tilde{\psi}_{f1}$ and the other representing thresholds that minimize L1 scores $\Tilde{\psi}_{l1}$ across all considered topologies. More specifically, we consider a shrinkage threshold selection using

\begin{equation} \label{threshold_rule_smith_2023_psi}
    \psi = \pi_1\Tilde{\psi}_{f1} + \pi_2\Tilde{\psi}_{l1},
\end{equation}\par

\noindent where $\pi_1 = \pi_2 = 0.5$ and from the networks considered here it is suggested that $\psi=0.12$. Figure \ref{fig:elastic_net_threshold_study} illustrates F1 scores across network topologies versus increasing threshold, $\psi$. Each subplot represents a model with thresholds denoted: 'dot-dash' for minimum L1 loss and 'dotted' for balanced F1 score/L1 loss optimization. 

%The parameter $\psi$ for this study is calculated to be 0.12, adhering to the methodology previously detailed. This calculation is founded on synthetic samples generated via the Markov Chain Monte Carlo (MCMC) sampling procedure described in Section \ref{sec:synthetic_examples}.

%%% thresh study plots

\subsection{Developing a block Gibbs sampler}\label{subsec:block_gibbs_sampler_for_bayesian_elastic_net}
The block Gibbs sampling procedures in \citep{wang_2012_efficient} and \citep{smith2022data} provide the foundational impetus for the sampler delineated in the current discussion. The posterior distribution associated with the hierarchical representation in \eqref{eq:bayes_BAE_posterior} can be expressed as

\begin{equation} \label{eq:hierarchical_BAE_target_dist}
    \begin{aligned}
        p(\mathbf{\Omega},\boldsymbol{\phi}|\mathbf{Y},\mu=0,\sigma,\lambda)
        &\propto
        \mathrm{det}\mathbf{\Omega}^{\frac{n}{2}}
        \exp\{-\mathrm{trace}(\frac{1}{2}\mathbf{S}\mathbf{\Omega})\}\\
        &\times 
        \prod_{i<j}\bigg\{\phi_{ij}^{-\frac{1}{2}}\exp(-\frac{\omega_{ij}^2}{2\phi_{ij}})\exp(-\frac{\lambda^2}{2}\phi_{ij})\bigg\}
        \prod_{i=1}^{p}\bigg\{\exp(-\frac{\lambda}{2}\omega_{ii})\bigg\}\\
        &\times 
        \prod_{i<j}\bigg\{\frac{1}{\sigma\sqrt{2\pi}}\exp\left(-\frac{\omega_{ij}^2}{2\sigma^2}\right)\bigg\}
    \prod_{i=1}^{p}\bigg\{\frac{\sqrt{2}}{\sigma\sqrt{\pi}}\exp\left(-\frac{\omega_{ii}^2}{2\sigma^2}\right)\bigg\}\mathds{1}_{\mathbf{\Omega}\in \mathbb{M}^{+}}.
    \end{aligned}
\end{equation}

\noindent It is important to highlight that the positive definiteness restriction is specifically relevant to the elements of $\mathbf{\Omega}$. \par

The block Gibbs sampler for \eqref{eq:hierarchical_BAE_target_dist} is detailed in this section. It employs a single-column-and-row update strategy per iteration. Without loss of generality, consider the final column and row of $\boldsymbol{\Omega}$. We first define $\boldsymbol{\Upsilon}$ as a symmetric $p\times p$ matrix with a zero main diagonal and $\boldsymbol{\phi}$ populating the upper off-diagonal entries. Next, the matrices $\boldsymbol{\Omega}$, $\boldsymbol{S}$, and $\boldsymbol{\Upsilon}$ are partitioned accordingly

\begin{equation} \label{eq:partition_gibbs}
    \boldsymbol{\Omega}=
        \begin{pmatrix}
         \boldsymbol{\Omega_{11}} & \boldsymbol{\omega_{12}}\\
         \boldsymbol{\omega_{21}} & \omega_{22}
        \end{pmatrix},
    \:\:\:\:
    \boldsymbol{S} = 
        \begin{pmatrix}
         \boldsymbol{S_{11}} & \boldsymbol{s_{12}}\\
         \boldsymbol{s_{21}} & s_{22}
        \end{pmatrix},
    \:\:\:\:
    \boldsymbol{\Upsilon} = 
        \begin{pmatrix}
         \boldsymbol{\Upsilon_{11}} & \boldsymbol{\phi_{12}}\\
         \boldsymbol{\phi_{21}} & 0
        \end{pmatrix}.
\end{equation}

\noindent Recall that the $\mathrm{det}\mathbf{\Omega}^{\frac{n}{2}}$ can be represented as
\begin{equation*}
    \begin{aligned}
        \mathrm{det}\mathbf{\Omega}^{\frac{n}{2}} = (\omega_{22}-\boldsymbol{\omega}_{21}\boldsymbol{\Omega}^{-1}_{11}\boldsymbol{\omega}_{12})^{\frac{n}{2}}\mathrm{det}\mathbf{\Omega_{11}}^{\frac{n}{2}}
        &\propto
        (\omega_{22}-\boldsymbol{\omega}_{21}\boldsymbol{\Omega}^{-1}_{11}\boldsymbol{\omega}_{12})^{\frac{n}{2}},
    \end{aligned}
\end{equation*}
since we are only interested in the last column and row. Similarly, 
\begin{equation*} 
    \mathrm{trace}(\frac{1}{2}\mathbf{S}\mathbf{\Omega}) \propto -\frac{1}{2}(2\boldsymbol{s_{21}}\boldsymbol{\omega}_{12} + s_{22}\omega_{22}).
\end{equation*}

%%% Paragraph 3
\noindent Notice that the conditional distribution of the last column in \eqref{eq:hierarchical_BAE_target_dist} can be presented as

\begin{equation} \label{eq:block_gibbs_conditional_dist_last_column}
    \begin{aligned}
        p(\boldsymbol{\omega_{12}},\omega_{22}\:|\: \boldsymbol{\Omega}_{11},\boldsymbol{\Upsilon},\boldsymbol{S},\mu\:,\sigma,\:\lambda)
        &\propto 
        (\omega_{22}-\boldsymbol{\omega}_{21}\boldsymbol{\Omega}^{-1}_{11}\boldsymbol{\omega}_{12})^{\frac{n}{2}}\\
        &\times
        \exp\left[-\frac{1}{2}\{\boldsymbol{\omega_{21}}\boldsymbol{D}^{*}\boldsymbol{\omega_{12}}
        + 2\boldsymbol{s_{21}}\boldsymbol{\omega}_{12} +
        + \omega_{22}\left(s_{22} + \lambda + 1\right)\}\right],
    \end{aligned}
\end{equation}

\noindent Where $\boldsymbol{D}^{*}=\mathrm{diag}\{\left(\sigma^2+\phi\right)/\left(\sigma^2\phi\right)\}$. Consider the following change of variables

\begin{equation*}
    \begin{aligned}
        \boldsymbol{\beta} &=\boldsymbol{\omega}_{12}\\
        \gamma &= \omega_{22}-\boldsymbol{\omega}_{21}\boldsymbol{\Omega}^{-1}_{11}\boldsymbol{\omega}_{12}
    \end{aligned}
\end{equation*}

%%% Paragraph 5
\noindent with Jacobian independent of $(\boldsymbol{\beta},\gamma)$, yields the following conditional distribution

\begin{equation*}
    \begin{aligned}
        p(\gamma\:,\boldsymbol{\beta}\:|\: \boldsymbol{\Omega}_{11},\boldsymbol{\Upsilon},\boldsymbol{S},\mu\:,\sigma,\:\lambda)
    \propto
    &(\gamma)^{\frac{n}{2}}\exp(-\frac{s_{22}+\lambda +1}{2}\gamma)\\
    &\times
    \exp\left[-\frac{1}{2}(\boldsymbol{\beta}^{\top}\{\boldsymbol{\phi}^{-1} + (s_{22} +\lambda + 1)\boldsymbol{\Omega}_{11}^{-1} \}\boldsymbol{\beta}+2\boldsymbol{s}_{21}\boldsymbol{\beta})\right].
    \end{aligned}
\end{equation*}

\noindent It follows that
\begin{equation*}
    (\gamma\:,\boldsymbol{\beta}|\: \boldsymbol{\Omega}_{11},\boldsymbol{\Upsilon},\boldsymbol{S},\lambda) 
    \sim \mathrm{GA}(\frac{n}{2}+1,\frac{s_{22}+\lambda + 1}{2})
    \mathrm{N}(-\boldsymbol{C}\boldsymbol{s}_{21},\boldsymbol{C}),
\end{equation*}

%%% Paragraph 6
\noindent where $\boldsymbol{C}=\{(s_{22}+\lambda + 1)\boldsymbol{\Omega}_{11}+\boldsymbol{D}^{*}\}^{-1}$. Furthermore, the update of $\boldsymbol{\phi}$ can be performed by acknowledging that the conditional posterior distribution of the $1/\phi_{ij}$'s in \eqref{eq:hierarchical_BAE_target_dist} are independently inverse Gaussian (INV-GAU) with parameters $\lambda_{ij}^{'}=\lambda_{ij}^2$ and $\mu^{'}=\sqrt{(\lambda_{ij}^2/\omega_{ij}^2)}$ and density

\begin{equation*}
    p(x)=(\frac{\lambda_{ij}^{'}}{2x\pi^3})^\frac{1}{2}
    exp\bigg\{\frac{-\lambda_{ij}^{'}(x-\mu^{'})^2}{2(\mu^{'})^2x}\bigg\}, \:\:\:x\:>\:0.
\end{equation*} 

The block Gibbs sampler upholds the positive definiteness restriction on $\mathbf{\Omega}$. For a thorough understanding of this constraint, readers may refer to the in-depth exposition given for the Bayesian graphical lasso sampler. Conclusively, the block Gibbs sampler can be summarized as follows

\begin{algorithm}
	\caption{Block Gibbs sampler}\label{alg:block_gibbs_sampler_BAE} 
	
	\begin{algorithmic}[]
	\Require Initialise $\mathbf{\Omega}$.
		\For {$i=1,\ldots,p$}
		    \State 1) Partition $\mathbf{\Omega}$, $\mathbf{S}$ and $\mathbf{\Upsilon}$ as in \eqref{eq:partition_gibbs}.
		    \State 2) Sample $\gamma \sim \mathrm{GA}(n/2+1,(s_{22}+\lambda + 1)/2)$ and $\boldsymbol{\beta} \sim \mathrm{N_{p-1}}\left(-\boldsymbol{C}\boldsymbol{s}_{21},\boldsymbol{C}\right)$.
		    \State 3) Update $\boldsymbol{\omega}_{12} = \boldsymbol{\beta}$, $\boldsymbol{\omega}_{21} = \boldsymbol{\beta}^{\top}$ and $\omega_{22} = \gamma + \boldsymbol{\beta}^{\top}\boldsymbol{\Omega_{11}}^{-1}\boldsymbol{\beta}$.
		\EndFor
            \For {$i \neq j$}
		    \State Sample $\delta_{ij} \sim \mathrm{INV}\mathrm{-}\mathrm{GAU}(\mu^{'},\lambda^{'})$ where $\lambda_{ij}^{'}=\lambda_{ij}^2$ and $\mu^{'}=\sqrt{(\lambda_{ij}^2/\omega_{ij}^2)}$ and update $\tau_{ij}=1/\delta_{ij}$.
		\EndFor
	\end{algorithmic} 
\end{algorithm}

\newpage
\subsection{An adaptive extension}\label{subsec:bayesian_adaptive_graphical_BAE}
Revisiting the double exponential prior depicted in \eqref{eq:bayes_glasso_prior} and \eqref{eq:bayes_graphical_BAE_prior}, it's important to acknowledge its known restrictions. It has a tendency to over-shrink larger coefficients while not shrinking smaller ones enough. These tendencies have been extensively explored in the fields of regression and graphical modeling, as evidenced by works such as \citep{carvalho2010horseshoe, griffin2010inference, wang_2012_efficient}, sparking the creation of alternate priors. Mirroring the methodology of the Bayesian graphical lasso, the hierarchical form in Section \ref{subsec:the_elastic_net_prior} and the block Gibbs sampler in Section \ref{subsec:block_gibbs_sampler_for_bayesian_elastic_net} may be tailored to allow for adaptive shrinkage on the off-diagonal elements $\omega_{ij}, \: i<j$. To this end, we propose the naïve Bayesian adaptive graphical elastic net which is a collection of prior distributions for $\lambda_{ij}$ and $\tau_{ij} = 1/\sigma_{ij}^2$.

\begin{equation} \label{eq:bayesian_adaptive_elastic_net}
    \begin{aligned}    
            p\left(\mathbf{\Omega}\:|\mu=0,\{\tau_{ij},\: \lambda_{ij}\}_{i<j}\right)=&C_{\{\tau_{ij}, \lambda_{ij}\}_{i<j}}^{-1}\prod_{i<j}\bigg\{\mathrm{DE}(\omega_{ij}\:|\:\lambda_{ij})\bigg\}
        \prod_{i=1}^{p}\bigg\{\mathrm{EXP}(\omega_{ii}\:|\:\lambda_{ii})\bigg\}\\
        & \times\prod_{i<j}\bigg\{\mathrm{N}(\omega_{ij}\:|\mu,\tau_{ij})\bigg\}
    \prod_{i=1}^{p}\bigg\{\mathrm{TN}(\omega_{ii}\:|\mu,\tau_{ii})\bigg\}\mathds{1}_{\mathbf{\Omega}\in \mathbb{M}^{+}}\\
    p\left(\{\tau_{ij}\}_{i<j}\:|\{\tau_{ii}\}^{p}_{i=1},\lambda \right)& \propto C_{\{\tau_{ij}\}_{i<j}}\prod_{i<j}\mathrm{EXP}(r)\\
        p(\{\lambda_{ij}\}_{i<j}\:|\:\{\lambda_{ii}\}_{i=1}^p,\tau)& \propto
    C_{\{\lambda_{ij}\}_{i < j}}
    \prod_{i<j}\mathrm{EXP}(s).
    \end{aligned}
\end{equation}

\noindent where $C_{\{\tau_{ij}, \lambda_{ij}\}_{i<j}}^{-1} = C^{-1}_{\{\tau_{ij}\}_{i<j}} C^{-1}_{\{\lambda_{ij}\}_{i < j}}$ is the intractable normalising constant. In this structure, we employ exponential distributions to update the parameters $\lambda_{ij}$ and $\tau_{ij}$, rather than using separate gamma and inverse gamma distributions as in \eqref{eq:bayesian_adaptive_lasso_prior} and \eqref{eq:bayesian_adaptive_ridge_prior} respectively. This strategic choice aims to benefit from the strengths of the involved components, while avoiding an increase in the number of necessary hyperparameters. Conditional on $\boldsymbol{\Omega}$, the target distributions for $\lambda_{ij}$ and ${\tau_{ij}}$ are given by

\begin{equation}\label{eq:posterior_sig_ij}
    \lambda_{ij}\:|\:\mathbf{\Omega} \sim \mathrm{GA}(1,|\omega_{ij}|+s)
\end{equation}

\begin{equation}\label{eq:posterior_tau_ij}
    \tau_{ij}\:|\:\mathbf{\Omega} \sim \mathrm{GA}\left(\frac{3}{2},\frac{\omega_{ij}^2}{2} + r\right).
\end{equation}

\noindent Notice that the conditional expected values for $\lambda_{ij}$ and $\tau_{ij}$ are $1/(|\omega_{ij}|+s)$ and $3/(\omega_{ij}^2+2r)$ respectively implying that the shrinkage applied will tend to be inversely proportional to the magnitude of $\omega_{ij}$. 

\subsection{Computational insights}\label{subsec:computational_insights}

In this section, we are focused on scrutinizing the efficacy and scalability of the block Gibbs sampler, elaborated in Algorithm \ref{alg:block_gibbs_sampler_BAE}. For this analysis, the target precision matrices considered are defined in Section \ref{sec:synthetic_examples}. For each network topology considered, precision matrix is initialised at the identity matrix. Performance assessments are executed on Apple silicon (M1 Pro), utilizing macOS Ventura (version 13.4), with computations being carried out through the R programming language (version 4.2.3). \par

\begin{figure}

\centering
\includegraphics[scale=.5]{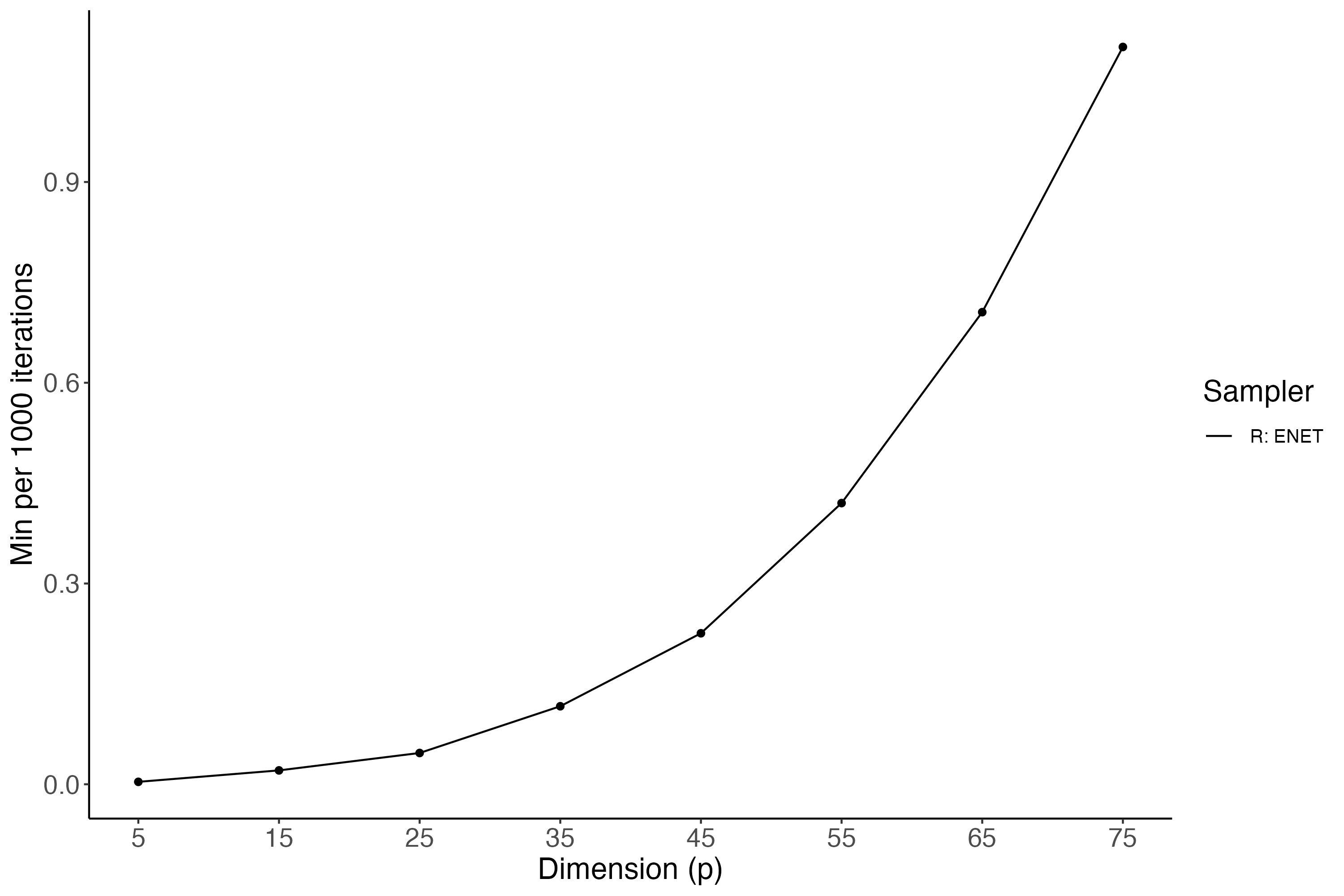}
\caption{Computational time comparisons as a function of $p$.}

\label{fig:computational_speed}
\end{figure}

For each network considered, Figure \ref{fig:computational_speed} displays the time it took the block Gibbs sampler to iteratively update $\mathbf{\Omega}$ 1000 times for increasing $p$. For example, the computation time required to perform 1000 iterations when $p=75$ is approximately 1 minute. Clearly the naïve Bayesian elastic net Gibbs sampler exhibits an exponential association between the computation time required to generate all of the entries of $\mathbf{\Omega}$ and $p$.  This is not surprising given that the sampler considered here builds upon the samplers of its building block constituents - whom also exhibit the same computational constraints for larger $p$. 

Convergence of the naïve Bayesian elastic net Gibbs sampler is assessed using the inefficiency factor $1+2\sum_{i=1}^{\infty}\eta(k)$, where $\eta(k)$ is the sample autocorrelation at lag $k$, as proposed by Kim (1998). The assessment procedure involves generating 3000 samples following a burn-in period of 5000 iterations and accommodating 300 lags. Median inefficiency factor is calculated for all elements of the precision matrix, $\mathbf{\Omega}$. This process is repeated 30 times, resulting in a median of the median inefficiency factors across all elements of $\mathbf{\Omega}$ of 0.9. This outcome suggests efficient mixing in the MCMC process.

\section{Synthetic examples}\label{sec:synthetic_examples}
The simulation study is designed to investigate the estimation of parameters and the determination of structure in synthetic DNs by employing the Bayesian adaptive graphical lasso, the Bayesian adaptive graphical ridge-type, and the naïve Bayesian adaptive graphical elastic net estimators (BAGL, BAGR, and BAE), respectively. For all simulations, the assumption is that the observations, $\mathbf{x}_1, \mathbf{x}_2, ... ,\mathbf{x}_{n_1}$ and $\mathbf{y}_1, \mathbf{y}_2, ... ,\mathbf{y}_{n_2}$ are generated from a Gaussian $N_p(0,\mathbf{\Sigma}_{1})$ and $N_p(0,\mathbf{\Sigma}_{2})$ respectively. The true DN is given by

\begin{equation*}
    \mathbf{\Delta}=\mathbf{\Sigma}_{2}^{-1} - \mathbf{\Sigma}_{1}^{-1},
\end{equation*}

%%% Paragraph
\noindent where the true precision matrices are $\mathbf{\Omega}_1=\mathbf{\Sigma}_{1}^{-1}$ and $\mathbf{\Omega}_2=\mathbf{\Sigma}_{2}^{-1}$. The components of the DNs, precision matrices, are estimated separately using the BAGL, BAGR and BAE estimators. For the BAGL estimator, the gamma prior parameters for $\lambda^{BAGL}_{ij}$ with $i < j$ are set to $r = 10^{-2}$ and $s = 10^{-6}$, as recommended by the original author, while $\lambda_{ii} = 1$. Subsequently, the parameters of the inverse gamma priors for $\tau^{BAGR}_{ij} = \sigma_{ij}^2$ with $i < j$ for the BAGR estimator are set to $a = 1$ and $b = 10^{-2}$. Lastly, from Section \ref{subsec:bayesian_adaptive_graphical_BAE}, the initial parameter choices for the BAE estimator are $r = 0.5$ for $\tau^{BAE}_{ij}$ and $s = 0.05$ for $\lambda^{BAE}_{ij}$. Note that $i = 1,\ldots,p$ applies to all priors. Table \ref{tab:syn_studies_models} presents the graphical structures examined in this study. These structures were specifically chosen to evaluate both sparse and non-sparse scenarios. To maintain the desired structure in the DN, its components are designed to adhere to the structure, with the first component being a scaled version of the second. The dimensions and sample sizes for each model are set as $p_1 = p_2 \in \{10, 30, 50\}$, with $n_1 = 10 * p_1$ and $n_2 = 10 * p_2$. The posterior distributions of each Bayesian $\mathbf{\Omega}$ estimate are derived using 10,000 Monte Carlo iterations, following an initial burn-in period of 5,000 iterations. The numerical performance evaluation of the DN estimation follows a methodology akin to that in \citep{smith2022empowering}, utilizing four loss functions as detailed in Table \ref{tab:syn_loss_functions}. Here, $p$ represents the dimension, while $\gamma_i$ denotes the $i^{th}$ eigenvalue. Eigenvalue-based loss functions offer valuable insights into the performance of the estimators within the eigenvalue spectrum. Tables \ref{tab:syn_loss_results_med} and \ref{tab:syn_loss_results_sd} display the median and standard error of L1, L2, EL1, and EL2, based on 50 replications. The best-performing result is highlighted in bold. The numerical loss results offer valuable insights into the strengths and weaknesses of the Bayesian estimators. Examining Tables \ref{tab:syn_loss_results_med} and \ref{tab:syn_loss_results_sd}, the BAGL estimator demonstrates exceptional performance for structures M2, M3, M4 and M6 as $p$ increases. Moreover, the BAGR estimator emerges as the evident frontrunner for structure M5 across all $p$ values and for M5 when. The BAE estimator, on the other hand, exhibits remarkable adaptability by incorporating a well-balanced blend of the numerical capabilities of both the BAGL and BAGR estimators. This is evident as the BAE estimator consistently ranks as either the best or the runner-up across all structures (with the exception of M5) and $p$ values for the L1 and L2 loss categories, as well as for the majority of the EL1 and EL2 loss categories.

A similar pattern is observed in the standard errors of the estimators, with the BAGL and BAE estimators demonstrating superiority here. This performance is not unexpected, given that all estimators employ a penalty parameter for each element of $\Omega$ when estimating the covariance and precision matrices, resulting in robust estimation. As demonstrated in Table \ref{tab:syn_perf_results}, the BAGL estimator favors sparsity, leading to reduced and more consistent standard errors.\par

Examining the graphical structure performance, Table \ref{tab:syn_perf_results} presents the median of sensitivity, specificity, precision, false negatives, F1 score, and balanced accuracy, based on 50 replications. According to the metrics defined in Table \ref{tab:syn_perf_functions}, values closer to one indicate superior classification performance, with the best-performing results highlighted in bold. In this context, TP, TN, FP, and FN represent the number of true positives, true negatives, false positives, and false negatives, respectively.

The BAGR estimator achieves the highest accuracy, F1 score, and sensitivity for M1 across all $p$ values, whilst the BAE and BAGL estimators dominate the specificity and precision. All estimators exhibit near-perfect to perfect sensitivity for M2-M4 and M6 across all $p$ values, respectively. The BAGL estimator attains the highest specificity, precision, F1 score, and accuracy for M2-M6 when $p=10$. The BAE estimator consistently ranks as either the best or the runner-up across all structures and $p$. For all estimators the standard error around the median are within 10\% of the median values for all performance results.

In summary, by incorporating a well-balanced blend of the numerical capabilities of both the BAGL and BAGR estimators, the BAE estimator consistently ranks among the top two performers in terms of loss metrics, graphical structure performance, and standard errors. This adaptability and robustness make the BAE estimator an ideal choice, providing reliable and accurate estimates in various situations, regardless of the sparsity level of the underlying structures.

%% WE SHOULD SHOW IN THE PHD THESIS HOW THE HYPERPARAMETERS OF BAE MOVE IT'S PERFORMANCE CLOSER TO THE BAGR OR BAGL

%% performance functions
\begin{table}[H]
\centering
\begin{tabular}{lll}
\hline                              

Measure & Performance function & Abbreviation  \\
\hline
Sensitivity & $\frac{\mathrm{TP}}{\mathrm{TP}+\mathrm{FN}}$ & SE \\
Specificity &  $\frac{\mathrm{TN}}{\mathrm{TN}+\mathrm{FP}}$ &  SP  \\
Precision & $\frac{\mathrm{TP}}{\mathrm{TP}+\mathrm{FP}}$ & PR \\
F1 score  & $\frac{2\mathrm{TP}}{2\mathrm{TP}+\mathrm{FP}+\mathrm{FN}}$ & F1  \\
Balanced accuracy  & $\frac{\mathrm{SE}+\mathrm{SP}}{2}$  & AC \\
\hline
\end{tabular}
\caption{Performance measures employed to evaluate the classification accuracy of the BAGL, BAGR and BAE estimates.}
\label{tab:syn_perf_functions}
\end{table}

\begin{table}[h]
\centering
\begin{tabular}{p{0.08\linewidth} | p{0.12\linewidth} | p{0.35 \linewidth}| p{0.35 \linewidth}}
\hline
\textbf{Model} & \textbf{Structure} & \textbf{Component 1} & \textbf{Component 2} \\ 
\hline
M1     & AR(1) & $\omega_{ij}=0.7^{|i-j|}$. & $\omega_{ij}=0.75^{|i-j|}$      \\
      % &          &            \\
M2     & AR(2)    & $\omega_{ii}=0.1, \omega_{i,i-1}= \omega_{i-1,i}= 0.05$ and $\omega_{i,i-2}= \omega_{i-2,i} = 0.025$. & $\omega_{ii}=1, \omega_{i,i-1}= \omega_{i-1,i}= 0.5$ and $\omega_{i,i-2}= \omega_{i-2,i} = 0.25$.       \\

M3    & Scale Free  & A scale-free model scaled to 0.5 of $\Omega_2$. & A scale-free model generated using B-A algorithm via the 'huge' R package.      \\
      % &          &            \\
      % &          &            \\
M4     & Band  & 
$\omega_{ii}=1,\omega_{ij}=0.2$ for $1 \leq i \neq j \leq p/2$, $\omega_{ij}=0.5$ for $p/2+1 \leq i \neq j \leq p$ and $\omega_{ij}=0$ otherwise. &
$\omega_{ii}=1,\omega_{ij}=0.7$ for $1 \leq i \neq j \leq p/2$, $\omega_{ij}=0.9$ for $p/2+1 \leq i \neq j \leq p$ and $\omega_{ij}=0$ otherwise.\\

M5  & Cluster  & 
$\omega_{ii}=1,\omega_{ij}=0.5$ for $1 \leq i \neq j \leq p/2$, $\omega_{ij}=0.5$ for $p/2+1 \leq i \neq j \leq p$ and $\omega_{ij}=0$ otherwise. & 
$\omega_{ii}=2,\omega_{ij}=1$ for $1 \leq i \neq j \leq p/2$, $\omega_{ij}=1$ for $p/2+1 \leq i \neq j \leq p$ and $\omega_{ij}=0$ otherwise.      \\

M6 & Circle &
$\omega_{ii}=2, \omega_{i,i-1}= \omega_{i-1,i}= 1$ and $\omega_{1,p}= \omega_{p,1} = 0.45$. &
$\omega_{ii}=4, \omega_{i,i-1}= \omega_{i-1,i}= 2$ and $\omega_{1,p}= \omega_{p,1} = 0.95$.  \\ 

\hline
\end{tabular}
\caption{Precision matrix structures and element compositions employed in the synthetic examples.}
\label{tab:syn_studies_models}
\end{table}

% \begin{itemize}
% %  \item \emph{Model 1}: A diagonal matrix with elements randomly sampled from Uniform distribution.
% %   \begin{itemize}
% %     \item \emph{Component 1}: $\sigma_{ij}\sim \mathrm{UNI}(1,1.25)$.
% %     \item \emph{Component 2}: $\sigma_{ij}\sim \mathrm{UNI}(1,1.2)$.
% %   \end{itemize}
%   \item \emph{Model 1}: A diagonal model with $\omega_{ii} \sim \mathrm{U(1,\:1.25)}$. Where, $\mathrm{U(\cdot)}$ denotes a uniform distribution.
%   \item \emph{Model 2}: An AR($1$) model with $\omega_{ij}=0.99^{|i-j|}$.
%   \item \emph{Model 3}: An AR($p-3$) model with $\omega_{ii}=3$ and $\omega_{p-3,p}=\omega_{p,p-3}=0.1$ and the sequential decay from $\omega_{ii}$ to $\omega_{p-3,i}$ and $\omega_{i,p-3}$ is $2.9/p$ in magnitude.
%   \item \emph{Model 4}: A cluster model containing two disjoint groups. $\omega_{ii}=2,\omega_{ij}=1$ for $1 \leq i \neq j \leq p/2$, $\omega_{ij}=1$ for $p/2+1 \leq i \neq j \leq p$ and $\omega_{ij}=0$ otherwise.
%   \item \emph{Model 5}: A cluster model containing two disjoint groups. $\omega_{ii}=2,\omega_{ij}=1$ for $1 \leq i \neq j \leq p/5$, $\omega_{ij}=1$ for $p/5+1 \leq i \neq j \leq p$ and $\omega_{ij}=0$ otherwise.
%   \item \emph{Model 6}: A full model with $\omega_{ii}=2$ and $\omega_{ij}=1$.
% \end{itemize}

%% loss functions
\begin{table}
\centering
\begin{tabular}{lll}
\hline                              

Measure                               & Loss function & Abbreviation  \\
\hline
Matrix $L_1$-norm                     & $\Arrowvert \hat{\mathbf{\Omega}} -  \mathbf{\Omega}\Arrowvert_1 = max_{1 \leq j \leq p}\sum_{i=1}^p|\hat{\Omega}_{ij}-\Omega_{ij}|$              & L1              \\

Frobenius loss                        &  $\Arrowvert \hat{\mathbf{\Omega}} -  \mathbf{\Omega}\Arrowvert_F$, where $\Arrowvert \boldsymbol{A}\Arrowvert_F^2=\mathrm{trace}(\boldsymbol{A}\boldsymbol{A}^{\top})$             &  L2             \\

$L_1$ eigenvalue loss                 & $\sum_{i=1}^p|\hat{\gamma}_i-\gamma_i|/p$              &   EL1             \\
$L_2$ eigenvalue loss                  & $\sum_{i=1}^p(\hat{\gamma}_i-\gamma_i)^2/p$              & EL2              \\
\hline
\end{tabular}
\caption{Loss functions utilized in the synthetic studies to evaluate the numerical accuracy of the BAGL, BAGR, and BAE estimates.}
\label{tab:syn_loss_functions}
\end{table}

%% Loss table
\begin{sidewaystable}[!ht]
%\begin{table}
%\centering
%\noindent\setlength\tabcolsep{4pt}%
\centering
\noindent\setlength\tabcolsep{4pt}%
\scriptsize
\begin{tabular}{lllllllllllllllllllll} 
\hline
\multicolumn{1}{c}{} & \multicolumn{3}{c}{\bf M1}  & \multicolumn{3}{c}{\bf M2} & \multicolumn{3}{c}{\bf M3} & \multicolumn{3}{c}{\bf M4} & \multicolumn{3}{c}{\bf M5}  & \multicolumn{3}{c}{\bf M6} \\ 
\cmidrule(lr){2-4}\cmidrule(lr){5-7}\cmidrule(lr){8-10}\cmidrule(lr){11-13}\cmidrule(lr){14-16}\cmidrule(lr){17-19}
\multicolumn{1}{c}{} & \multicolumn{1}{c}{BAE} & \multicolumn{1}{c}{BAGL} & \multicolumn{1}{c}{BAGR} & \multicolumn{1}{c}{BAE} & \multicolumn{1}{c}{BAGL} & \multicolumn{1}{c}{BAGR} & \multicolumn{1}{c}{BAE} & \multicolumn{1}{c}{BAGL} & \multicolumn{1}{c}{BAGR} & \multicolumn{1}{c}{BAE} & \multicolumn{1}{c}{BAGL} & \multicolumn{1}{c}{BAGR} & \multicolumn{1}{c}{BAE} & \multicolumn{1}{c}{BAGL} & \multicolumn{1}{c}{BAGR} & \multicolumn{1}{c}{BAE} & \multicolumn{1}{c}{BAGL} & \multicolumn{1}{c}{BAGR}  \\ 
\hline
\multicolumn{21}{c}{p=10} 
\\
\\
L1 & \textbf{2.08} & 2.2 & 3.04 
    & \textbf{0.89} & 0.94 & 1.66 
    & 3.03 & \textbf{2.1} & 4.02 
    & 1.18 & \textbf{0.94} & 2.31 
    & \textbf{2.69} & 2.98 & 2.92 
    & 3.26 & \textbf{2.2} & 5.41 \\
    
L2 & 3.54 & \textbf{3.47} & 3.71 
    & \textbf{0.97} & 0.98 & 1.42 
    & 2.4 & \textbf{2.17} & 3.7 
    & 1.32 & \textbf{1.22} & 2.23 
    & 3 & 2.84 & \textbf{2.64} 
    & 4.32 & \textbf{2.55} & 4.75 \\
    
EL1 & 1.07 & \textbf{1.03} & \textbf{1.03} 
    & \textbf{0.12} & \textbf{0.12} & 0.21 
    & \textbf{0.26} & \textbf{0.26} & 0.57 
    & \textbf{0.2} & \textbf{0.2} & 0.32 
    & 0.54 & \textbf{0.45} & 0.48 
    & 0.85 & \textbf{0.33} & 0.64 \\
    
EL2 & 1.19 & \textbf{1.1} & 1.28 
    & \textbf{0.02} & 0.03 & 0.08 
    & 0.15 & \textbf{0.09} & 0.55 
    & \textbf{0.05} & 0.06 & 0.2 
    & 0.58 & 0.42 & \textbf{0.35} 
    & 1.13 & \textbf{0.19} & 0.73 \\
\\
\multicolumn{21}{c}{p=30}  
\\
\\

L1 & \textbf{2.82} & 2.91 & 4.27 
    & 0.57 & \textbf{0.52} & 1.48 
    & 2.7 & \textbf{1.83} & 4.34 
    & 0.87 & \textbf{0.76} & 2.08 
    & 8.75 & 7.79 & \textbf{4.42} 
    & 2.59 & \textbf{1.57} & 5.59 \\

L2 & \textbf{6.01} & 6.09 & 6.37 
    & 0.92 & \textbf{0.8} & 1.98 
    & 2.39 & \textbf{2.1} & 4.32 
    & 1.46 & \textbf{1.4} & 2.72 
    & 11.3 & 10.43 & \textbf{3.88} 
    & 5.23 & \textbf{2.56} & 6.92 \\
    
EL1 & 1.05 & 1.06 & \textbf{1.02} 
    & 0.06 & \textbf{0.04} & 0.2 
    & \textbf{0.12} & \textbf{0.12} & 0.4 
    & \textbf{0.1} & \textbf{0.1} & 0.22 
    & 0.78 & 0.65 & \textbf{0.38} 
    & 0.65 & \textbf{0.12} & 0.52 \\
    
EL2 & \textbf{1.14} & 1.15 & 1.24 
    & \textbf{0.01} & \textbf{0.01} & 0.08 
    & \textbf{0.03} & \textbf{0.03} & 0.28 
    & \textbf{0.02} & \textbf{0.02} & 0.09 
    & 2.99 & 3.25 & \textbf{0.28} 
    & 0.64 & \textbf{0.04} & 0.68 \\
 
\\
\multicolumn{21}{c}{p=50}                                                      \\&&&&&&&&&&&&&&&&&&&&\\

L1 & \textbf{2.8} & 2.91 & 3.89 
    & 0.54 & \textbf{0.43} & 1.23 
    & 2.32 & \textbf{1.74} & 4.25 
    & 0.67 & \textbf{0.58} & 1.52 
    & 13.61 & 12.76 & \textbf{5.49} 
    & 2.34 & \textbf{1.32} & 4.26 \\
    
L2 & \textbf{7.68} & 7.7 & 7.87 
    & 1.03 & \textbf{0.75} & 2.22 
    & 2.3 & \textbf{1.98} & 4.02 
    & 1.47 & \textbf{1.32} & 2.6 
    & 18.43 & 17.57 & \textbf{4.87} 
    & 5.84 & \textbf{2.57} & 6.45 \\
    
EL1 & \textbf{1.05} & \textbf{1.05} & 1.01 
    & 0.08 & \textbf{0.03} & 0.2 
    & 0.1 & \textbf{0.08} & 0.29 
    & 0.08 & \textbf{0.06} & 0.19 
    & 0.78 & 0.61 & \textbf{0.35} 
    & 0.57 & \textbf{0.08} & 0.41 \\
    
EL2 & 1.13 & \textbf{1.12} & 1.14 
    & \textbf{0.01} & \textbf{0.01} & 0.07 
    & 0.02 & \textbf{0.01} & 0.17 
    & \textbf{0.01} & \textbf{0.01} & 0.06 
    & 5.59 & 5.89 & \textbf{0.28} 
    & 0.48 & \textbf{0.02} & 0.38 \\ 
        
\hline
\end{tabular}
\caption{Summary of L1, L2, EL1, and EL2 loss values for all precision matrix structures in Table \ref{tab:syn_studies_models}. The median loss values reported in this table are derived from 50 replications for each of the BAGL, BAGR, and BAE estimators. The best-performing values are highlighted in bold.}
%\begin{table}
%\end{table}
\label{tab:syn_loss_results_med}
\end{sidewaystable}

\clearpage

%% SD table
\begin{sidewaystable}[!ht]
%\begin{table}
%\centering
%\noindent\setlength\tabcolsep{4pt}%
\centering
\noindent\setlength\tabcolsep{4pt}%
\scriptsize
\begin{tabular}{lllllllllllllllllllll} 
\hline
\multicolumn{1}{c}{} & \multicolumn{3}{c}{\bf M1}  & \multicolumn{3}{c}{\bf M2} & \multicolumn{3}{c}{\bf M3} & \multicolumn{3}{c}{\bf M4} & \multicolumn{3}{c}{\bf M5}  & \multicolumn{3}{c}{\bf M6} \\ 
\cmidrule(lr){2-4}\cmidrule(lr){5-7}\cmidrule(lr){8-10}\cmidrule(lr){11-13}\cmidrule(lr){14-16}\cmidrule(lr){17-19}
\multicolumn{1}{c}{} & \multicolumn{1}{c}{BAE} & \multicolumn{1}{c}{BAGL} & \multicolumn{1}{c}{BAGR} & \multicolumn{1}{c}{BAE} & \multicolumn{1}{c}{BAGL} & \multicolumn{1}{c}{BAGR} & \multicolumn{1}{c}{BAE} & \multicolumn{1}{c}{BAGL} & \multicolumn{1}{c}{BAGR} & \multicolumn{1}{c}{BAE} & \multicolumn{1}{c}{BAGL} & \multicolumn{1}{c}{BAGR} & \multicolumn{1}{c}{BAE} & \multicolumn{1}{c}{BAGL} & \multicolumn{1}{c}{BAGR} & \multicolumn{1}{c}{BAE} & \multicolumn{1}{c}{BAGL} & \multicolumn{1}{c}{BAGR}  \\ 
\hline
\multicolumn{21}{c}{p=10} 
\\
\\
L1 & \textbf{0.29} & 0.37 & 0.88 
    & \textbf{0.23} & 0.3 & 0.55 
    & \textbf{0.69} & 0.81 & 1.15 
    & \textbf{0.27} & 0.28 & 0.72 
    & \textbf{0.36} & 0.86 & 1.03 
    & \textbf{0.39} & 0.56 & 1.39 \\
    
L2 & \textbf{0.29} & 0.35 & 0.64 
    & \textbf{0.17} & 0.25 & 0.32 
    & \textbf{0.41} & \textbf{0.41} & 0.62 
    & \textbf{0.2} & 0.24 & 0.41 
    & \textbf{0.35} & 0.85 & 0.72 
    & \textbf{0.41} & 0.52 & 0.94 \\
    
EL1 & \textbf{0.09} & 0.11 & 0.13 
    & \textbf{0.03} & 0.04 & 0.07 
    & 0.1 & \textbf{0.09} & 0.14 
    & \textbf{0.05} & \textbf{0.05} & 0.09 
    & \textbf{0.06} & 0.11 & 0.12 
    & 0.13 & \textbf{0.12} & 0.19 \\
    
EL2 & \textbf{0.19} & 0.24 & 0.49 
    & \textbf{0.01} & 0.02 & 0.08 
    & \textbf{0.08} & 0.1 & 0.39 
    & \textbf{0.02} & 0.03 & 0.14 
    & \textbf{0.12} & 0.2 & 0.46 
    & 0.32 & \textbf{0.12} & 0.72\\ 
\\
\multicolumn{21}{c}{p=30}  
\\
\\
L1 & \textbf{0.19} & 0.22 & 0.51 
    & \textbf{0.08} & 0.09 & 0.34 
    & 0.97 & \textbf{0.57} & 1.24 
    & 0.15 & \textbf{0.12} & 0.48 
    & 0.25 & \textbf{0.12} & 1.39 
    & \textbf{0.35} & 0.39 & 1.16 \\
    
L2 & \textbf{0.14} & 0.16 & 0.28 
    & \textbf{0.09} & 0.09 & 0.2 
    & 0.39 & \textbf{0.26} & 0.6 
    & \textbf{0.15} & \textbf{0.15} & 0.29 
    & 0.17 & \textbf{0.11} & 0.81 
    & \textbf{0.29} & 0.34 & 0.48 \\
    
EL1 & \textbf{0.02} & \textbf{0.02} & 0.03 
    & \textbf{0.01} & \textbf{0.01} & 0.02 
    & \textbf{0.04} & \textbf{0.04} & 0.08 
    & \textbf{0.02} & \textbf{0.02} & 0.04 
    & \textbf{0.01} & 0.02 & 0.07 
    & 0.05 & \textbf{0.03} & 0.08 \\
    
EL2 & \textbf{0.05} & 0.06 & 0.09 
    & \textbf{0.01} & \textbf{0.01} & 0.02 
    & \textbf{0.02} & \textbf{0.02} & 0.15 
    & \textbf{0.01} & \textbf{0.01} & 0.04 
    & 0.05 & \textbf{0.04} & 0.23 
    & 0.1 & \textbf{0.02} & 0.17\\
    
\\
\multicolumn{21}{c}{p=50}                                                      \\&&&&&&&&&&&&&&&&&&&&\\
L1 & \textbf{0.23} & \textbf{0.23} & 0.4 
    & \textbf{0.11} & 0.1 & 0.19 
    & 1.58 & \textbf{0.76} & 2.06 
    & 0.14 & \textbf{0.11} & 0.27 
    & 0.16 & \textbf{0.06} & 1.24 
    & \textbf{0.23} & 0.31 & 0.96 \\
    
L2 & 0.13 & \textbf{0.12} & 0.18 
    & 0.1 & \textbf{0.08} & 0.16 
    & 0.42 & \textbf{0.24} & 0.47 
    & \textbf{0.11} & \textbf{0.11} & 0.19 
    & 0.14 & \textbf{0.03} & 0.69 
    & \textbf{0.29} & 0.32 & 0.67 \\
    
EL1 & 0.02 & \textbf{0.01} & 0.02 
    & \textbf{0.01} & \textbf{0.01} & 0.02 
    & \textbf{0.02} & \textbf{0.02} & 0.04 
    & \textbf{0.01} & \textbf{0.01} & 0.03 
    & \textbf{0.01} & \textbf{0.01} & 0.04 
    & 0.04 & \textbf{0.02} & 0.07 \\
    
EL2 & \textbf{0.03} & \textbf{0.03} & 0.05 
    & \textbf{0.01} & \textbf{0.01} & \textbf{0.01} 
    & \textbf{0.01} & \textbf{0.01} & 0.06 
    & \textbf{0.01} & \textbf{0.01} & 0.02 
    & \textbf{0.04} & 0.05 & 0.15
    & 0.06 & \textbf{0.01} & 0.12\\ 

\hline
\end{tabular}
\caption{Summary of the standard errors for L1, L2, EL1, and EL2 loss values associated with all precision matrix structures in Table \ref{tab:syn_studies_models}. The standard errors presented in this table are calculated based on 50 replications for the BAGL, BAGR, and BAE estimators. The best-performing values are highlighted in bold.}
%\begin{table}
%\end{table}
\label{tab:syn_loss_results_sd}
\end{sidewaystable}

\clearpage

%% Performance table
\begin{sidewaystable}[!ht]
%\begin{table}
%\centering
%\noindent\setlength\tabcolsep{4pt}%
\centering
\noindent\setlength\tabcolsep{4pt}%
\scriptsize
\begin{tabular}{lllllllllllllllllllll} 
\hline
\multicolumn{1}{c}{} & \multicolumn{3}{c}{\bf M1}  & \multicolumn{3}{c}{\bf M2} & \multicolumn{3}{c}{\bf M3} & \multicolumn{3}{c}{\bf M4} & \multicolumn{3}{c}{\bf M5}  & \multicolumn{3}{c}{\bf M6} \\ 
\cmidrule(lr){2-4}\cmidrule(lr){5-7}\cmidrule(lr){8-10}\cmidrule(lr){11-13}\cmidrule(lr){14-16}\cmidrule(lr){17-19}
\multicolumn{1}{c}{} & \multicolumn{1}{c}{BAE} & \multicolumn{1}{c}{BAGL} & \multicolumn{1}{c}{BAGR} & \multicolumn{1}{c}{BAE} & \multicolumn{1}{c}{BAGL} & \multicolumn{1}{c}{BAGR} & \multicolumn{1}{c}{BAE} & \multicolumn{1}{c}{BAGL} & \multicolumn{1}{c}{BAGR} & \multicolumn{1}{c}{BAE} & \multicolumn{1}{c}{BAGL} & \multicolumn{1}{c}{BAGR} & \multicolumn{1}{c}{BAE} & \multicolumn{1}{c}{BAGL} & \multicolumn{1}{c}{BAGR} & \multicolumn{1}{c}{BAE} & \multicolumn{1}{c}{BAGL} & \multicolumn{1}{c}{BAGR}  \\ 
\hline
\multicolumn{21}{c}{p=10} 
\\
\\

SE & 0.58 & 0.47 & \textbf{0.94} 
    & \textbf{1} & 0.95 & \textbf{1} 
    & \textbf{1} & \textbf{1} & \textbf{1} 
    & \textbf{1} & \textbf{1} & \textbf{1} 
    & \textbf{1} & \textbf{1} & \textbf{1} 
    & \textbf{1} & \textbf{1} & \textbf{1} \\
    
SP & \textbf{1} & \textbf{1} & \textbf{1} 
    & 0.81 & \textbf{0.96} & 0.61 
    & 0.89 & \textbf{0.97} & 0.58 
    & 0.92 & \textbf{0.97} & 0.56 
    & 0.92 & \textbf{1} & 0.57 
    & 0.93 & \textbf{1} & 0.61 \\
    
PR & \textbf{1} & \textbf{1} & \textbf{1} 
    & 0.79 & \textbf{0.95} & 0.66 
    & 0.77 & \textbf{0.93} & 0.47 
    & 0.82 & \textbf{0.93} & 0.46 
    & 0.92 & \textbf{1} & 0.69 
    & 0.85 & \textbf{1} & 0.51 \\
    
F1 & 0.73 & 0.64 & \textbf{0.97} 
    & 0.88 & \textbf{0.95} & 0.8 
    & 0.87 & \textbf{0.96} & 0.64 
    & 0.9 & \textbf{0.96} & 0.63 
    & 0.96 & \textbf{1} & 0.82 
    & 0.92 & \textbf{1} & 0.67 \\
    
AC & 0.79 & 0.74 & \textbf{0.97} 
    & 0.89 & \textbf{0.96} & 0.81 
    & 0.95 & \textbf{0.99} & 0.79 
    & 0.96 & \textbf{0.99} & 0.78 
    & 0.96 & \textbf{1} & 0.78 
    & 0.96 & \textbf{1} & 0.8 \\
\\
\multicolumn{21}{c}{p=30}  
\\
\\

SE & 0.27 & 0.24 & \textbf{0.54} 
    & \textbf{1} & \textbf{1} & \textbf{1} 
    & \textbf{1} & \textbf{1} & \textbf{1} 
    & \textbf{1} & \textbf{1} & \textbf{1} 
    & 0.96 & 0.27 & \textbf{1} 
    & \textbf{1} & \textbf{1} & \textbf{1} \\
    
SP & \textbf{1} & \textbf{1} & 0.95 
    & 0.98 & \textbf{1} & 0.91 
    & \textbf{0.99} & \textbf{0.99} & 0.9 
    & 0.99 & \textbf{1} & 0.91 
    & 0.99 & \textbf{1} & 0.94 
    & \textbf{1} & \textbf{1} & 0.89 \\
    
PR & \textbf{1} & \textbf{1} & \textbf{1} 
    & 0.89 & \textbf{1} & 0.68 
    & 0.9 & \textbf{0.94} & 0.53 
    & 0.9 & \textbf{0.98} & 0.53 
    & \textbf{0.99} & 0.98 & 0.94 
    & 0.96 & \textbf{1} & 0.51 \\
    
F1 & 0.42 & 0.39 & \textbf{0.7} 
    & 0.94 & \textbf{1} & 0.81 
    & 0.95 & \textbf{0.97} & 0.69 
    & 0.95 & \textbf{0.99} & 0.7 
    & \textbf{0.97} & 0.42 & \textbf{0.97} 
    & 0.98 & \textbf{1} & 0.67 \\
    
AC & 0.63 & 0.62 & \textbf{0.73} 
    & 0.99 & \textbf{1} & 0.96 
    & 0.99 & \textbf{1} & 0.95 
    & 0.99 & \textbf{1} & 0.95 
    & \textbf{0.97} & 0.63 & \textbf{0.97} 
    & \textbf{1} & \textbf{1} & 0.95 \\

\\
\multicolumn{21}{c}{p=50}                                                      \\&&&&&&&&&&&&&&&&&&&&\\
SE & 0.24 & 0.21 & \textbf{0.4} 
    & \textbf{1} & \textbf{1} & \textbf{1} 
    & \textbf{1} & \textbf{1} & \textbf{1} 
    & \textbf{1} & \textbf{1} & \textbf{1} 
    & 0.54 & 0.1 & \textbf{1} 
    & \textbf{1} & \textbf{1} & \textbf{1} \\
    
SP & \textbf{1} & \textbf{1} & 0.98 
    & \textbf{1} & \textbf{1} & 0.98 
    & \textbf{1} & \textbf{1} & 0.98 
    & \textbf{1} & \textbf{1} & 0.98 
    & \textbf{1} & \textbf{1} & 0.99 
    & \textbf{1} & \textbf{1} & 0.97 \\
    
PR & \textbf{1} & \textbf{1} & 0.98 
    & 0.96 & \textbf{1} & 0.83 
    & 0.96 & \textbf{0.99} & 0.75 
    & 0.97 & \textbf{0.99} & 0.75 
    & \textbf{0.99} & \textbf{0.99} & \textbf{0.99} 
    & 0.97 & \textbf{1} & 0.7 \\
    
F1 & 0.39 & 0.35 & \textbf{0.57} 
    & 0.98 & \textbf{1} & 0.91 
    & 0.98 & \textbf{0.99} & 0.85 
    & \textbf{0.99} & \textbf{0.99} & 0.86 
    & 0.7 & 0.19 & \textbf{1} 
    & 0.99 & \textbf{1} & 0.82 \\
    
AC & 0.62 & 0.61 & \textbf{0.69} 
    & \textbf{1} & \textbf{1} & 0.99 
    & \textbf{1} & \textbf{1} & 0.99 
    & \textbf{1} & \textbf{1} & 0.99 
    & 0.77 & 0.55 & \textbf{1} 
    & \textbf{1} & \textbf{1} & 0.99\\

\hline
\end{tabular}
\caption{Summary of the sensitivity, specificity, precision, F1 score, and balanced accuracy values for all precision matrix structures in Table \ref{tab:syn_studies_models}. The performance metrics in this table are derived from 50 replications for each of the BAGL, BAGR, and BAE estimators. The best-performing values are highlighted in bold.}
%\begin{table}
%\end{table}
\label{tab:syn_perf_results}
\end{sidewaystable}

\clearpage

\clearpage

\section{Application study}\label{sec:application_study}
In this section, we apply the Bayesian adaptive elastic net estimator to diverse real-world datasets encompassing oncology, nephrology, and enology. We scrutinize genomic alterations in cancer studies, investigate pathological mechanisms within kidney disease datasets, and utilise BAE in a wine study, not for classification, but to discern associations between physicochemical properties of wine varieties. These applications highlight the BAE estimator's capacity to navigate complex, multidimensional problems across distinct scientific domains.

\subsection{SEER breast cancer}\label{subsec:seer_breast_cancer_study}
The dataset utilised originates from the Surveillance, Epidemiology, and End Results (SEER) Program, managed by the National Cancer Institute (NCI), and is based on the November 2017 update \citep{seer_breast_cancer_data}. It specifically pertains to female patients diagnosed with infiltrating duct and lobular carcinoma breast cancer (identified by SEER primary cites recode NOS histology codes 8522/3) within the 2006-2010 timeframe. Data selection procedures ensured exclusion of patients with indeterminable tumor size, unknown status of regional Lymph Nodes (LNs), unconfirmed number of positive regional LNs, and those with survival duration less than a month. As a result, the final dataset encompasses data from 4024 patients.  

The aforementioned dataset is split into two distinct groups according to the patients' mortality status - those who survived and those who unfortunately succumbed to the disease. The application of the BAE estimator is conducted individually on each subset of data. In order to ascertain the differential network, the network derived from the cohort of patients who survived is subtracted from the network of those who passed away. This differential network facilitates the understanding of the distinctive interactions and relationships present in the data of each patient group.

\begin{figure}[H]
\vspace*{\fill} % For vertical centering
\begin{subfigure}{\linewidth}
  \centering
  \includegraphics[width=0.75\linewidth]{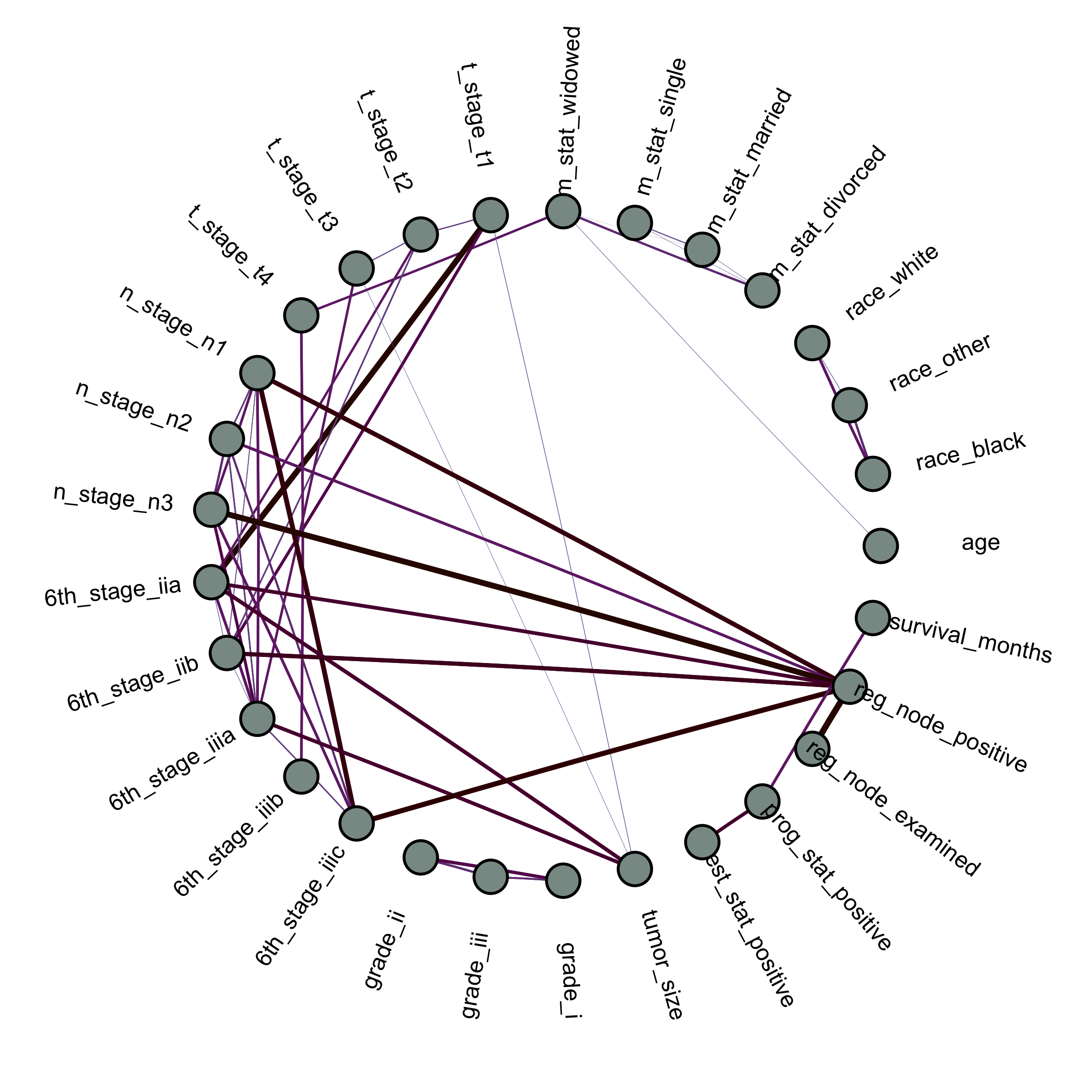}
  \vspace{-3em} % reduce vertical space
      \caption{Differential network between deceased and living patients.}
\end{subfigure}

\vspace{-0.15cm} % add some whitespace after the first figure

\begin{subfigure}[b]{0.45\linewidth}
  \centering
  \includegraphics[width=1.1\linewidth]{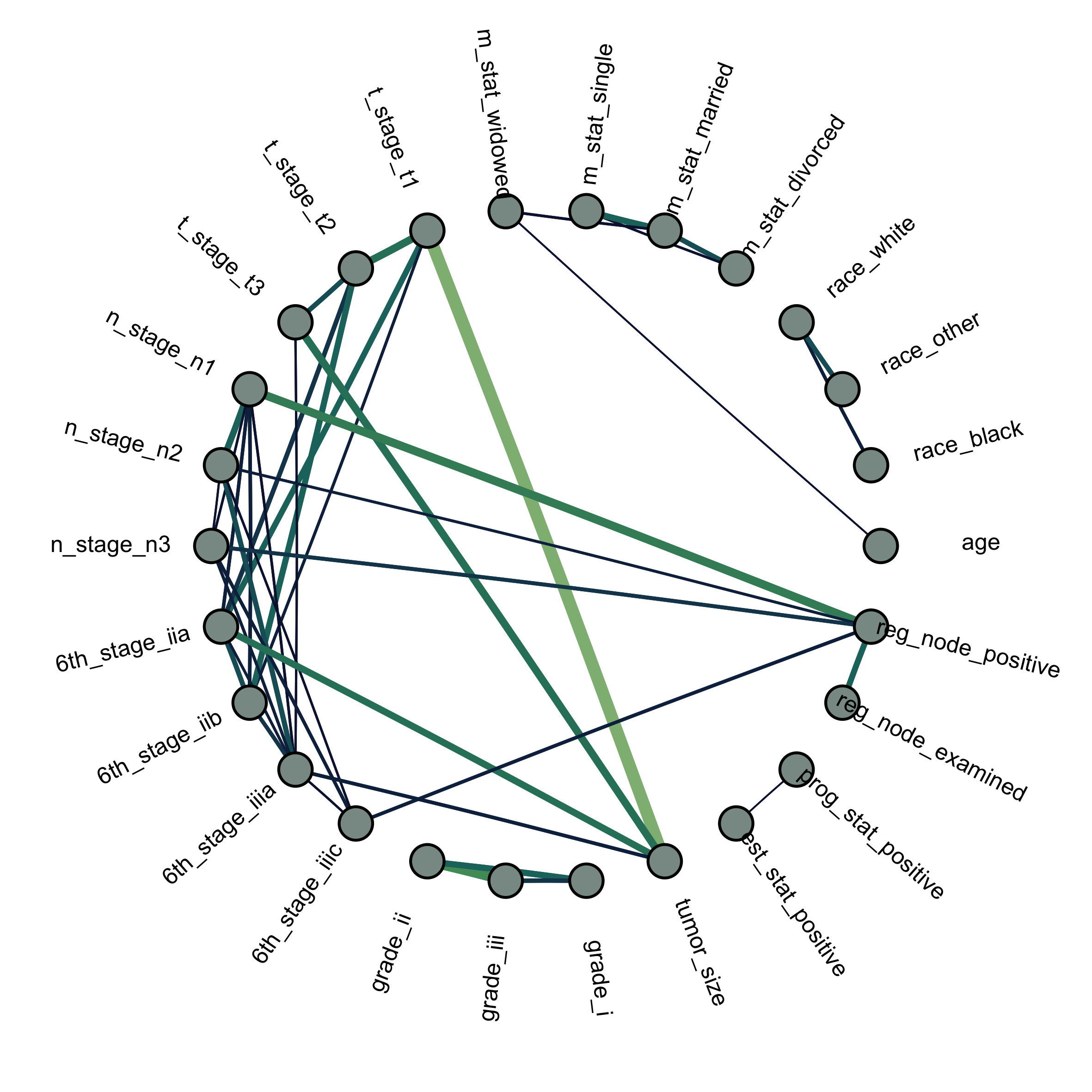}
  \vspace{-4em} % reduce vertical space
  \caption{Network for patients still alive.}
  \label{fig:blah1}
\end{subfigure}
\hfill
\begin{subfigure}[b]{0.45\linewidth}
  \centering
  \includegraphics[width=1.1\linewidth]{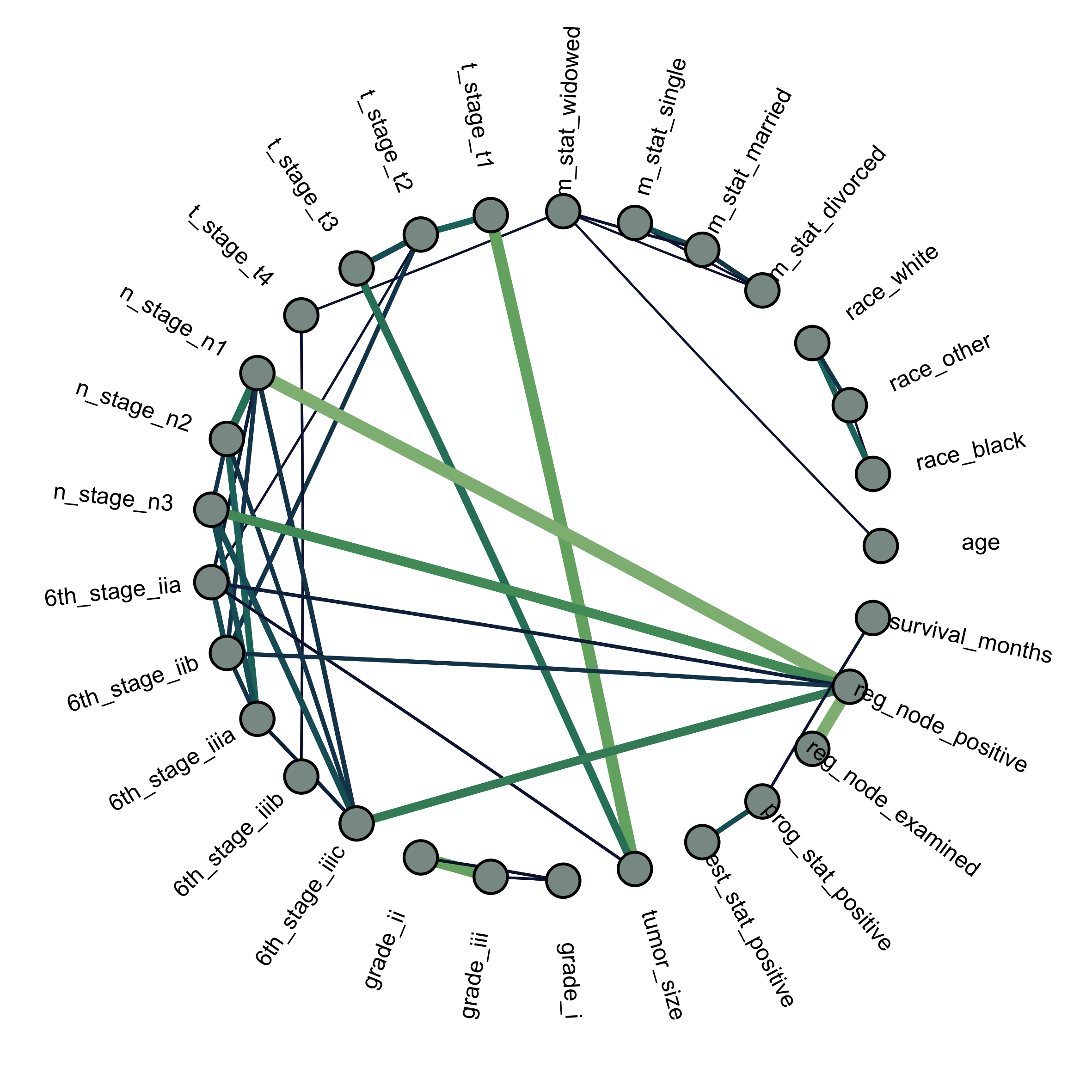}
  \vspace{-4em} % reduce vertical space
  \caption{Network for deceased patients.}
  \label{fig:blah2}
\end{subfigure}
\vspace*{\fill} % For vertical centering
\caption{Network estimations generated using the BAE estimator for the breast cancer dataset. (a) Differential network showcasing the distinction between the networks of deceased and surviving patients. (b) Network for the group of patients who are still alive. (c) Network for the group of deceased patients.}
\label{fig:seer_cancer_networks}
\end{figure}

Through examination of the differential network in Figure \ref{fig:seer_cancer_networks}, several noteworthy observations can be derived from the interconnections among various biomarker features. 
Firstly, a few expected clusters emerge between both cohorts, such as the ethnicity, marital status and the interaction between the Adjusted AJCC 6th ed. T, N, M,  stages \citep{frederick2002ajcc}. Moreover, deceased patients appear to exhibit stronger associations between the presence of metastases in the lymph nodes (regional node positive) and the degree to which the tumor has spread to axillary or mammary lymph nodes (N stage 3 and 6th stage IIIC). Secondly, there appears to be an association between a positive progesterone receptor status and the survival duration post-diagnosis in breast cancer patients \citep{li2020clinicopathological}. Thirdly, the presence of the tumor in T4 stage is only present in within the deceased cohort and is associated with the 6th stage IIIB classification signifying that the tumor has advanced into the chest wall, skin, or both, or it is classified as inflammatory breast cancer. In surviving patients, a stronger correlation is observed between smaller tumor size (T1) and the Stage IIA classification. This supports the findings in \citep{verschraegen2005modeling}, where there is functional relationship between tumor size and mortality. These findings underscore the intricate interconnections among various factors in the complex landscape of breast cancer, and highlight the potential of differential network analysis in elucidating these relationships.

\subsection{Chronic kidney disease}\label{subsec:chronic_kidney_disease_study}

The Chronic Kidney Disease (CKD) dataset under study  from \citep{misc_chronic_kidney_disease_336} was obtained from a hospital in Tamil Nadu, India, over a span of two months concluding in July 2015. This dataset consists of patient records, encompassing blood test results and various crucial medical measurements. Out of the 400 patient records, 250 are associated with individuals diagnosed with CKD, while the remaining 150 correspond to non-CKD patients.

This dataset incorporates a wide range of patient-specific metrics such as age, blood pressure, urine specific gravity, albumin and sugar levels in urine, red blood cell count, pus cell count and its clumps in urine, the presence of bacteria in urine, random blood glucose levels, blood urea, serum creatinine, sodium and potassium levels, hemoglobin, packed cell volume, white blood cell count, and red blood cell count. Furthermore, it documents the presence or absence of hypertension, diabetes mellitus, coronary artery disease, and anemia in patients. It also records patient details on appetite and pedal edema.

Each patient is categorised according to their CKD status. Similar to the breast cancer dataset, the BAE estimator is applied to this CKD dataset to gain a more profound understanding of the interrelationships between the aforementioned parameters and the cohorts. The differential network is derived by calculating the difference between the networks of patients without CKD and those diagnosed with the disease.

\begin{figure}[H]
\vspace*{\fill} % For vertical centering
\begin{subfigure}{\linewidth}
  \centering
  \includegraphics[width=0.7\linewidth]{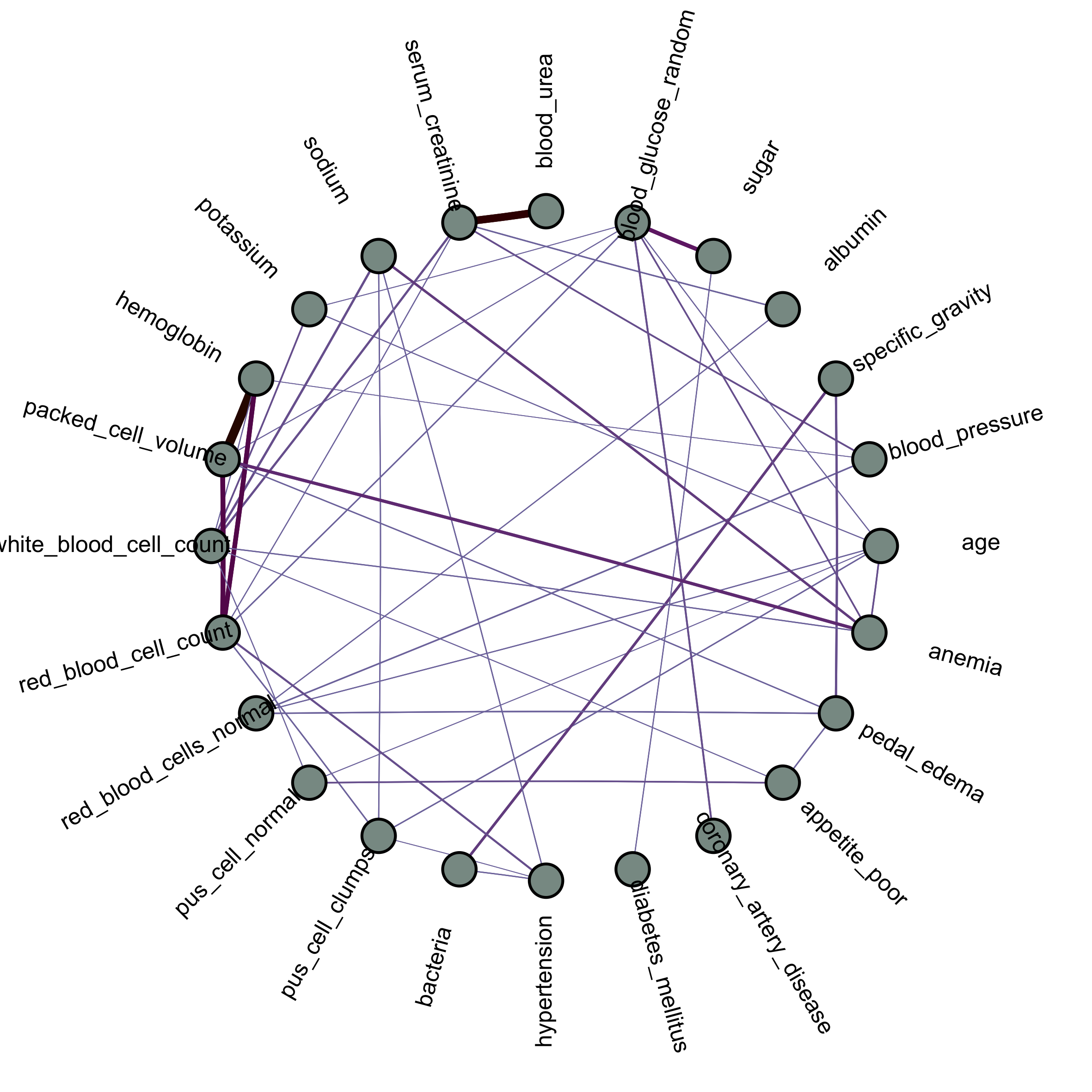}
  \vspace{-2em} % reduce vertical space
      \caption{Differential network between patients without and with CKD.}
\end{subfigure}

\vspace{-0.15cm} % add some whitespace after the first figure

\begin{subfigure}[b]{0.45\linewidth}
  \centering
  \includegraphics[width=1\linewidth]{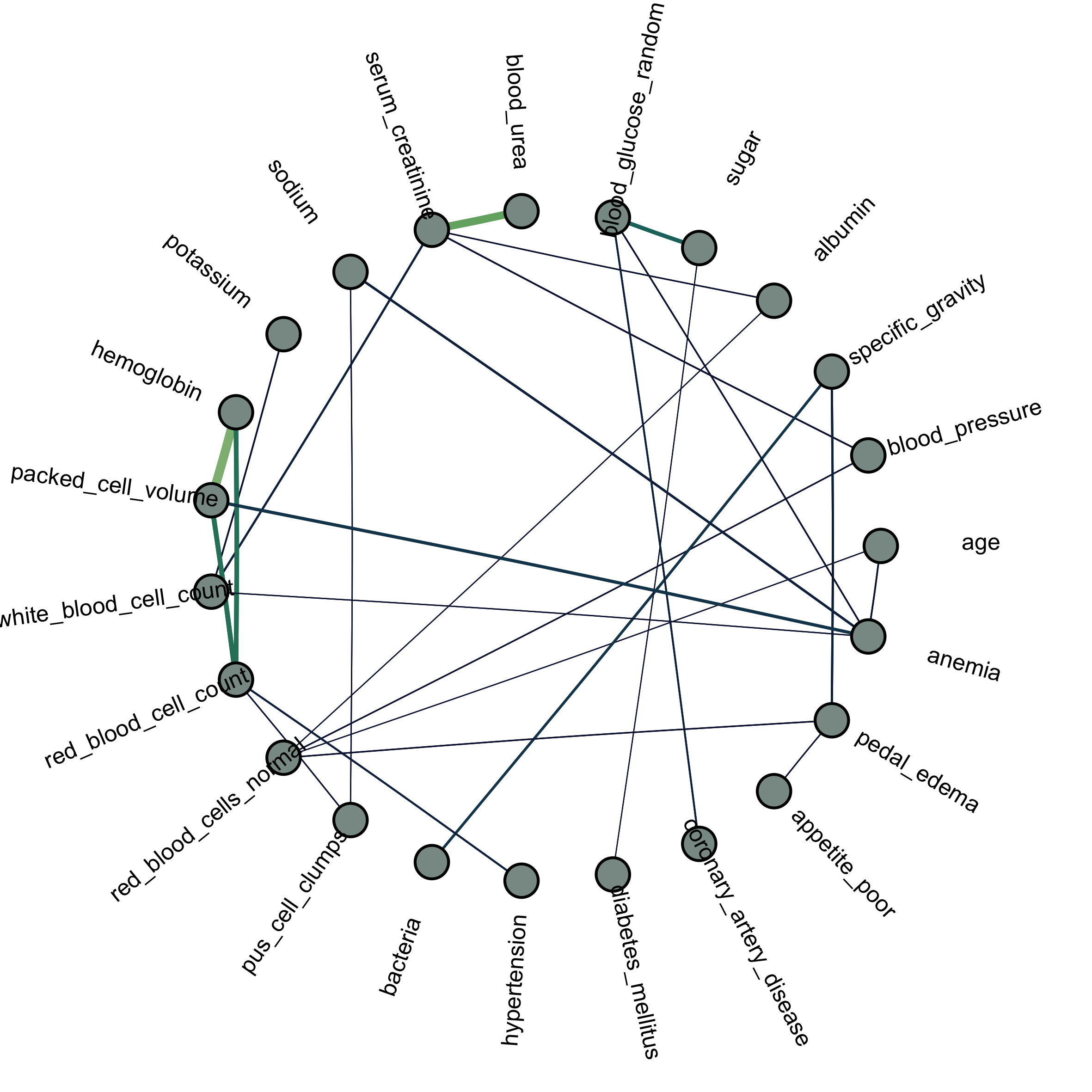}
  \vspace{-3em} % reduce vertical space
  \caption{Network for patients with CKD.}
  \label{fig:blah1}
\end{subfigure}
\hfill
\begin{subfigure}[b]{0.45\linewidth}
  \centering
  \includegraphics[width=1\linewidth]{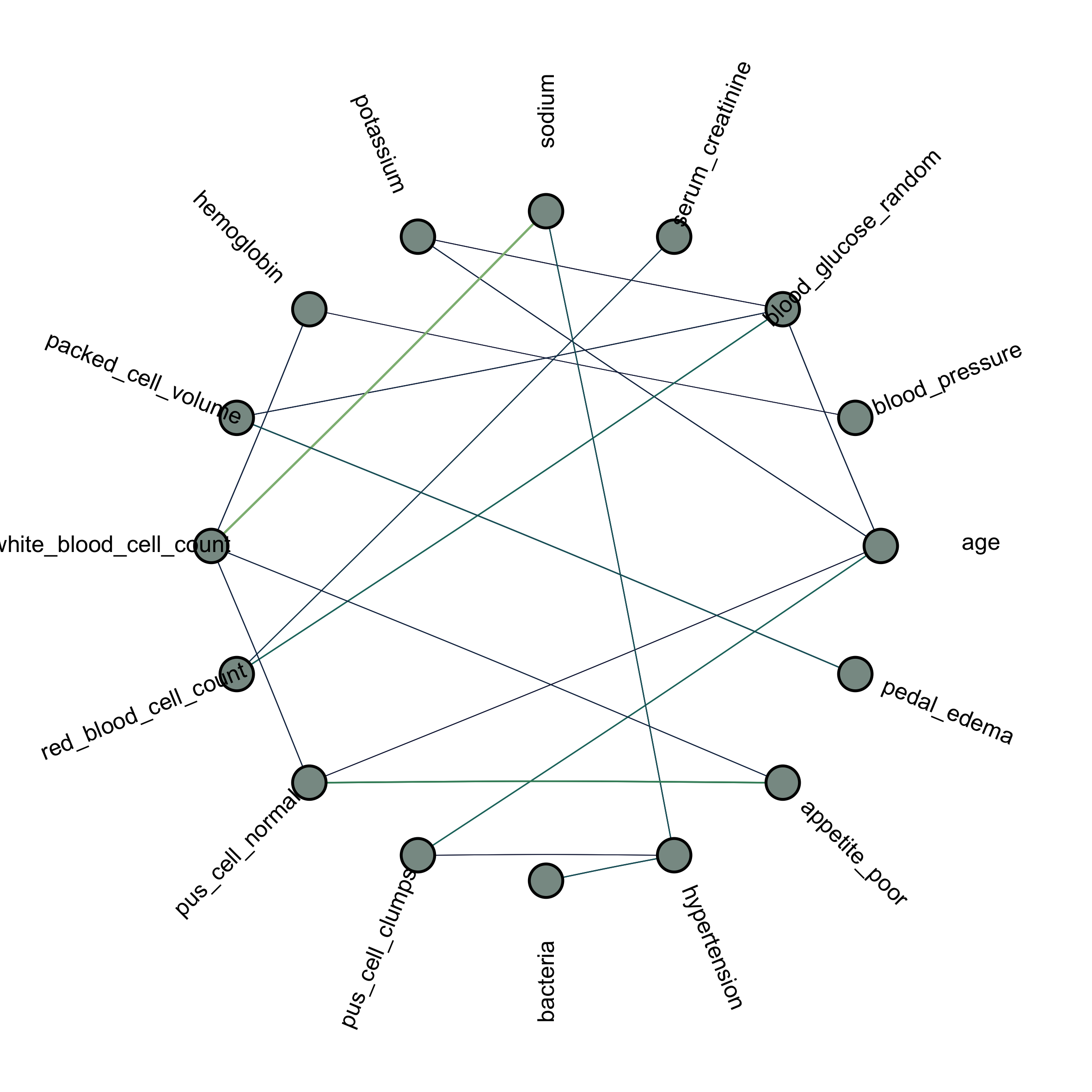}
  \vspace{-3em} % reduce vertical space
  \caption{Network for patients without CKD.}
  \label{fig:blah2}
\end{subfigure}
\vspace*{\fill} % For vertical centering
\caption{Visual representation of the networks derived from the Chronic Kidney Disease (CKD) study. Subfigure (a) represents the differential network, calculated as the difference between the networks of patients without from those with CKD. Subfigures (b) and (c) represent the network of patients diagnosed with CKD and those without CKD, respectively}
\label{fig:chronic_kidney_disease_networks}
\end{figure}

In examining the differential network in Figure \ref{fig:chronic_kidney_disease_networks}, several pertinent relationships emerge. Firstly, a relationship is observed between packed cell volume (PCV) levels, hemoglobin (Hb) and anemia suggesting a potential hematological interaction within the CKD population \citep{khanam2013relationship}. Additionally, a marked association is identified between blood urea and creatinine levels in patients with CKD, similar results were obtained in \citep{yadav2014evaluation}. A distinct correlation emerges between bacterial presence and urine specific gravity. Elevated bacterial counts and or higher specific gravity potentially suggest urinary tract infection (UTI) or CKD, an assertion resonating with \citep{nainggolan2021diagnostic}. Finally, notice the sparser network structure for patients without CKD and the absence of any concerning metrics. 

%These relationships uncovered by the differential network analysis underscore the complexity of CKD and its myriad associations.

\subsection{Italian wine cultivar }\label{subsec:wine_data_study}

This application involves an analysis of the well studied Italian wine dataset \citep{misc_wine_109}. This dataset, features wines derived from three distinct cultivars. It encapsulates 13 attributes, each providing unique insights into the chemical composition of the wines. These attributes include Alcohol, Malic Acid, Ash, Alkalinity of Ash, Magnesium, Total Phenols, Flavanoids, Nonflavanoid Phenols, Proanthocyanins, Color Intensity, Hue, OD280/OD315 of diluted wines, and Proline.

In our study, we employ the BAE estimator to dissect this dataset further and analyze the differential networks between class 1 and class 2, class 1 and class 3, and finally, class 2 and class 3. The aim is to understand the nuances and differentiations in these wine classes better, potentially unveiling unique characteristics and associations within and between these classes which may aid in the inferential deductions in the typical classification studies.

When examining the differential networks in Figures \ref{fig:bayes_dn_wine_class12} - \ref{fig:bayes_dn_wine_class23} between class 1 and class 2 wines, a couple of noteworthy associations are discernible. In contrast to class 2, Class 1 wines also exhibit a distinct link between color intensity and total phenols. Additionally, the hue in class 2 it is primarily associated with malic acid levels. In class 2 wines, flavanoids also demonstrate a strong link with the 0D280/0D315 of diluted wines levels.

In the differential network between class 1 and class 3 wines, the color intensity in class 3 wines appears to be potentially associated with proanthocyanins levels and the hue seems to be linked with magnesium levels. Additionally, the flavanoids in class 3 wines exhibit a pronounced association with magnesium levels.

Upon comparing class 2 and class 3 wines, the hue of class 2 wines shows an association with malic acids. For class 3 wines, ash content demonstrates an association with total phenols measurements. Notably, nonflavanoid phenols in class 3 wines show a stronger association with flavanoids.

These findings provide insights into the distinct chemometric profiles of the different wine classes. By understanding these underlying associations, we could potentially improve the inferential capabilities of machine learning models used for classification purposes.

\begin{table}[H]
\centering
\begin{tabular}{l l l}
\hline
\bf Bayesian DN & \bf First precision & \bf Second precision \\ 
\begin{subfigure}[c]{0.28\textwidth}
      \includegraphics[width=\textwidth]{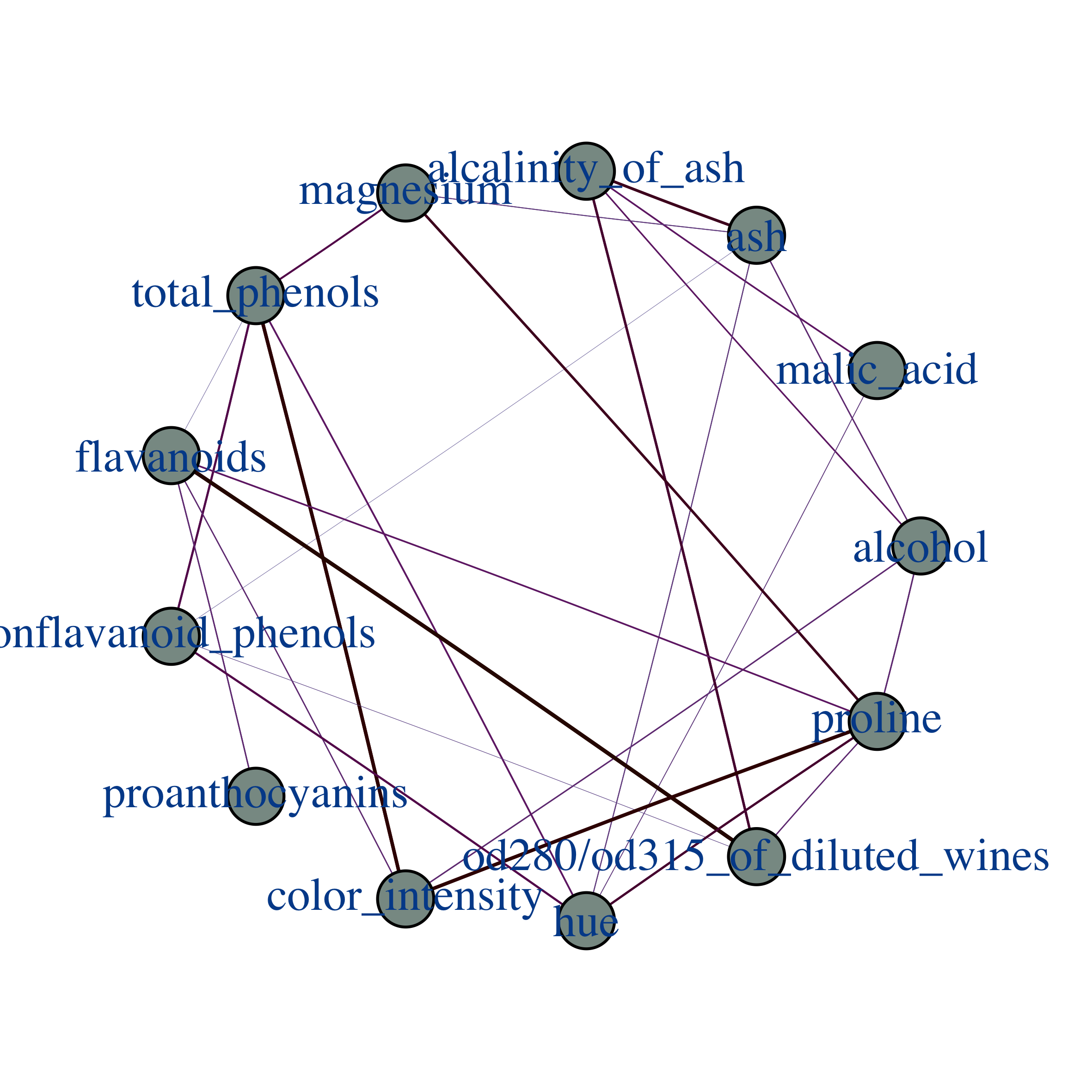}
         \caption{DN between class 1 and 2 wine networks.}
         \label{fig:bayes_dn_wine_class12}
    \end{subfigure}&
\begin{subfigure}[c]{0.28\textwidth}
      \includegraphics[width=\textwidth]{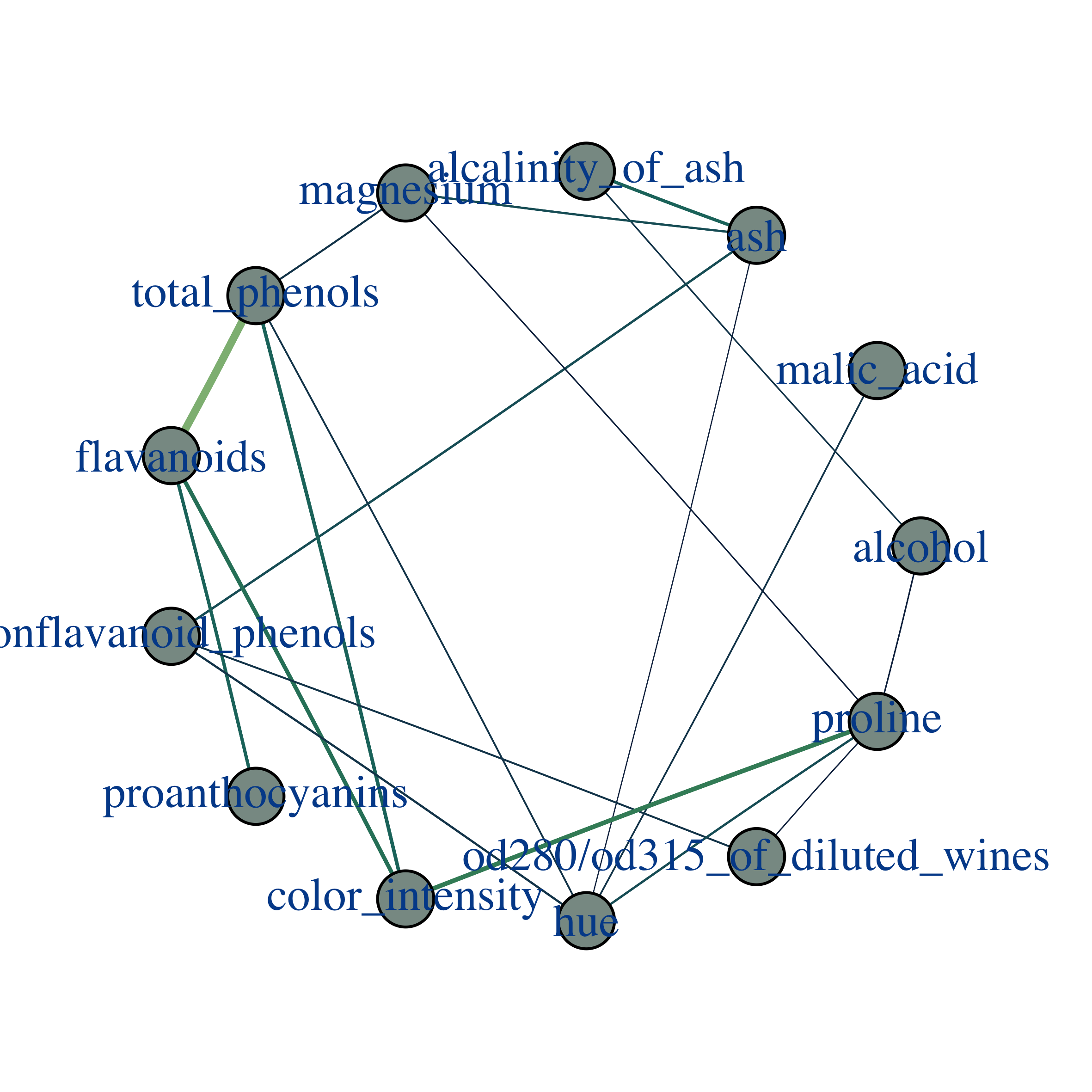}
         \caption{Class 1 wine precision matrix network.}
         \label{fig:class1_wine}
    \end{subfigure}&    
\begin{subfigure}[c]{0.28\textwidth}
      \includegraphics[width=\textwidth]{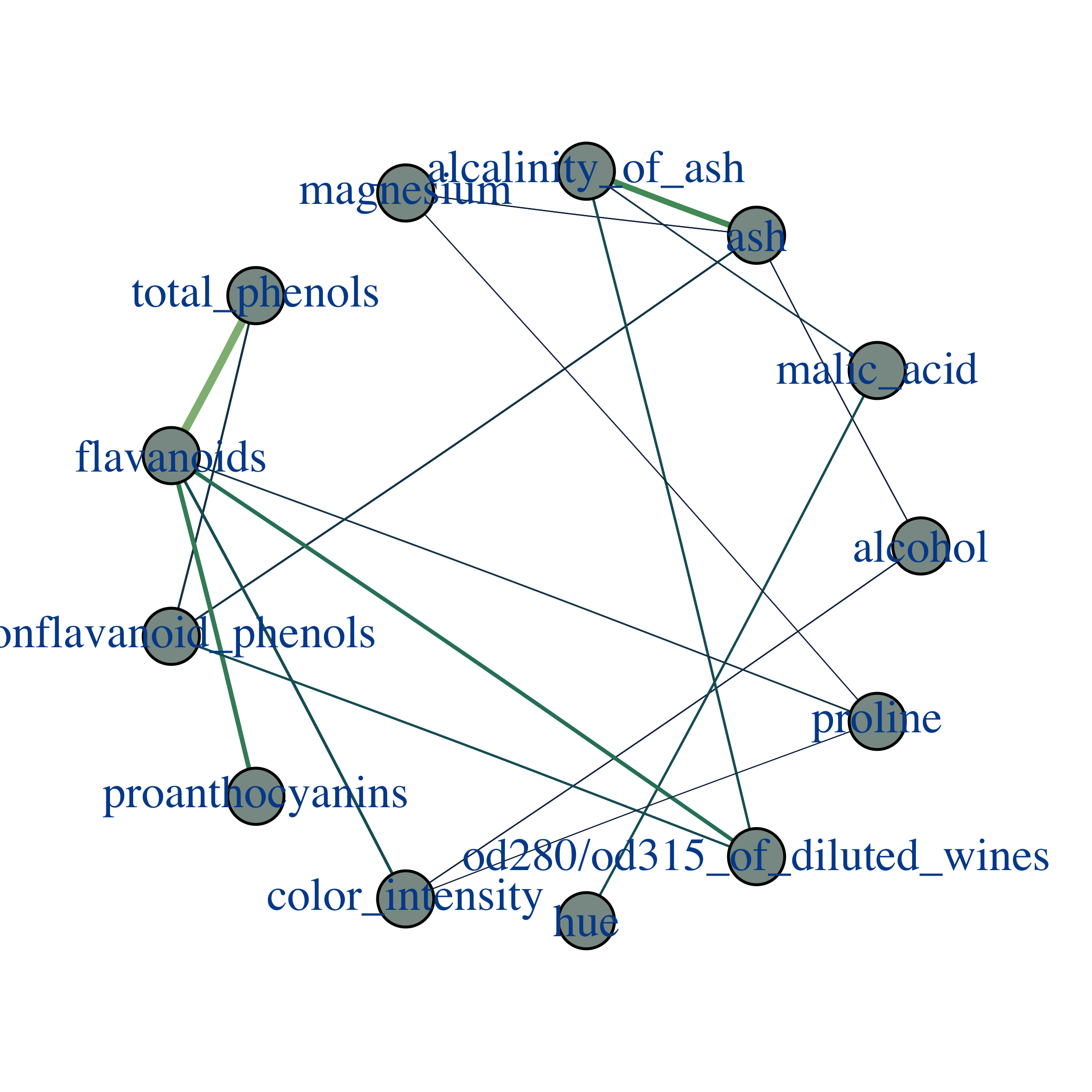}
         \caption{Class 2 wine precision matrix network.}
         \label{fig:class2_wine}
    \end{subfigure}\\
\hline
\begin{subfigure}[c]{0.28\textwidth}
      \includegraphics[width=\textwidth]{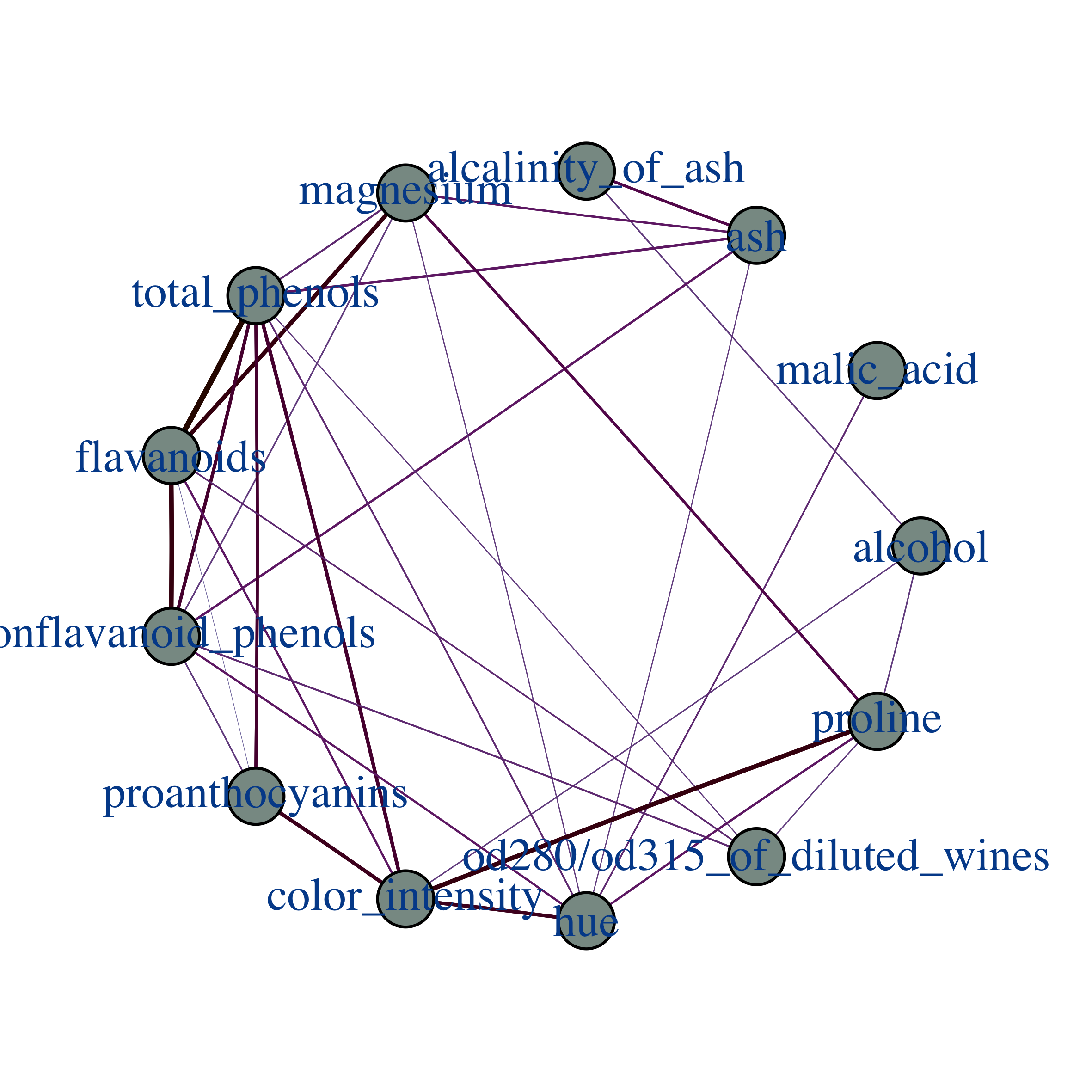}
         \caption{DN between class 1 and 3 wine networks.}
         \label{fig:bayes_dn_wine_class13}
    \end{subfigure}&    
\begin{subfigure}[c]{0.28\textwidth}
      \includegraphics[width=\textwidth]{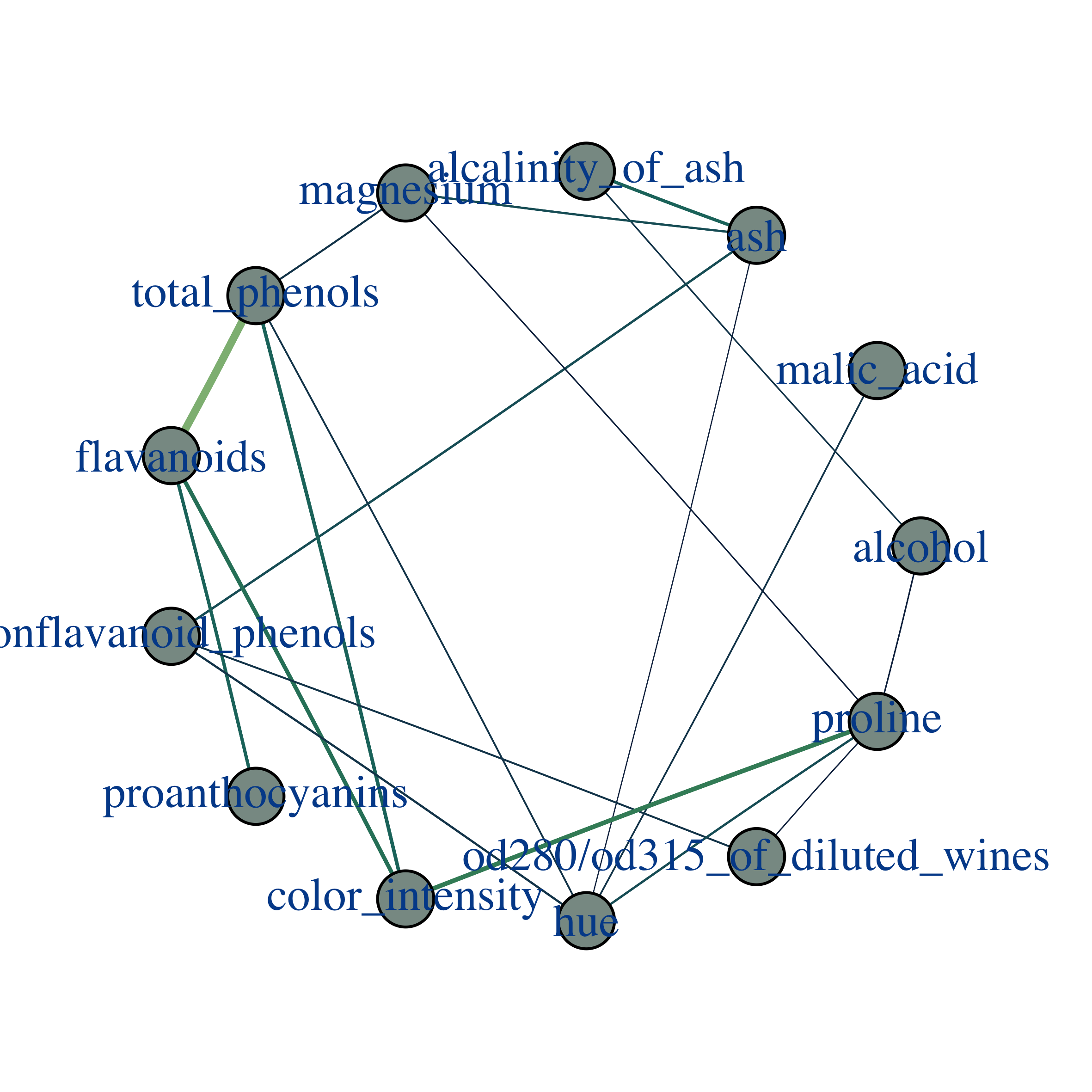}
         \caption{Class 1 wine precision matrix network.}
    \end{subfigure}&    
\begin{subfigure}[c]{0.28\textwidth}
      \includegraphics[width=\textwidth]{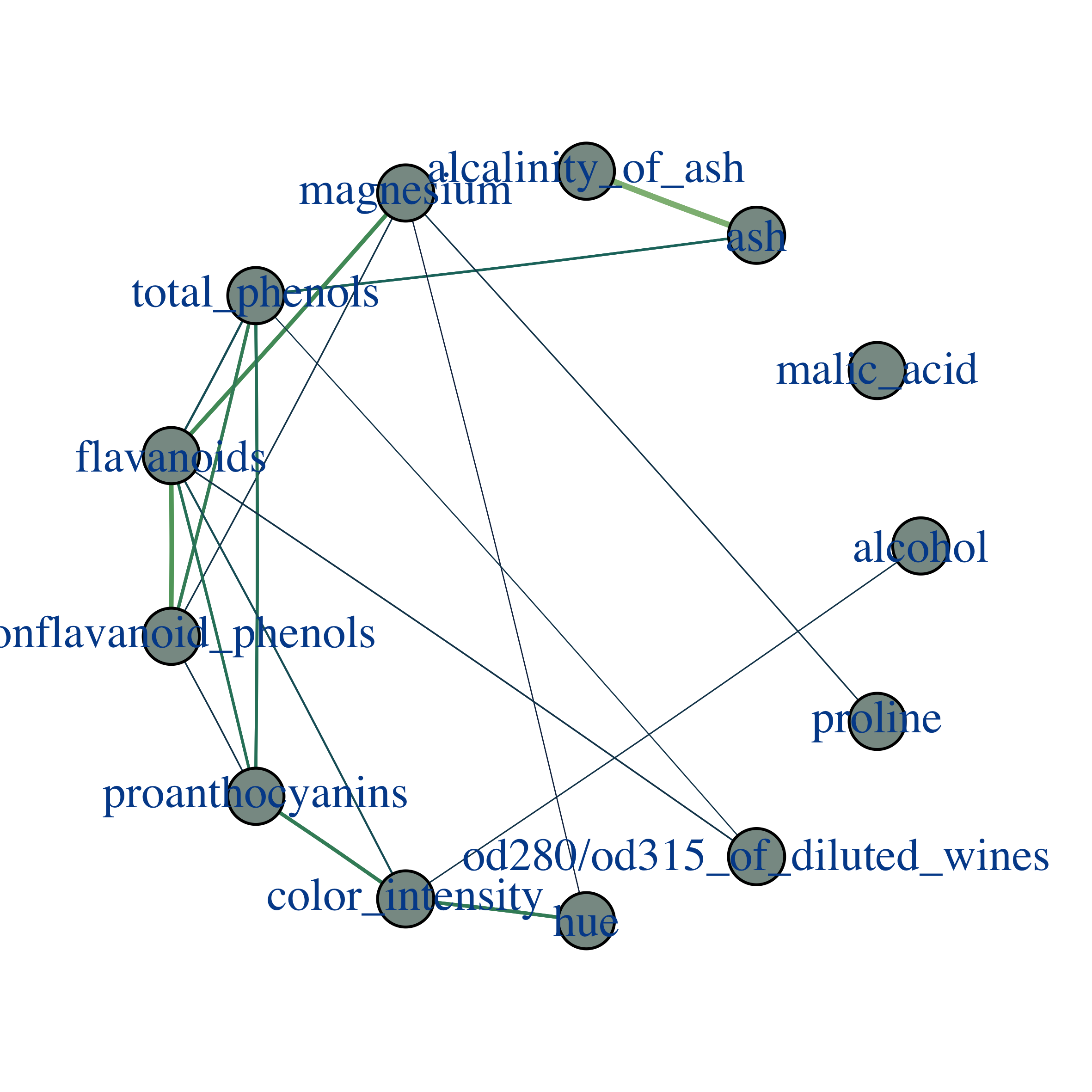}
         \caption{Class 3 wine precision matrix network.}
    \end{subfigure}\\
\hline
\begin{subfigure}[c]{0.28\textwidth}
      \includegraphics[width=\textwidth]{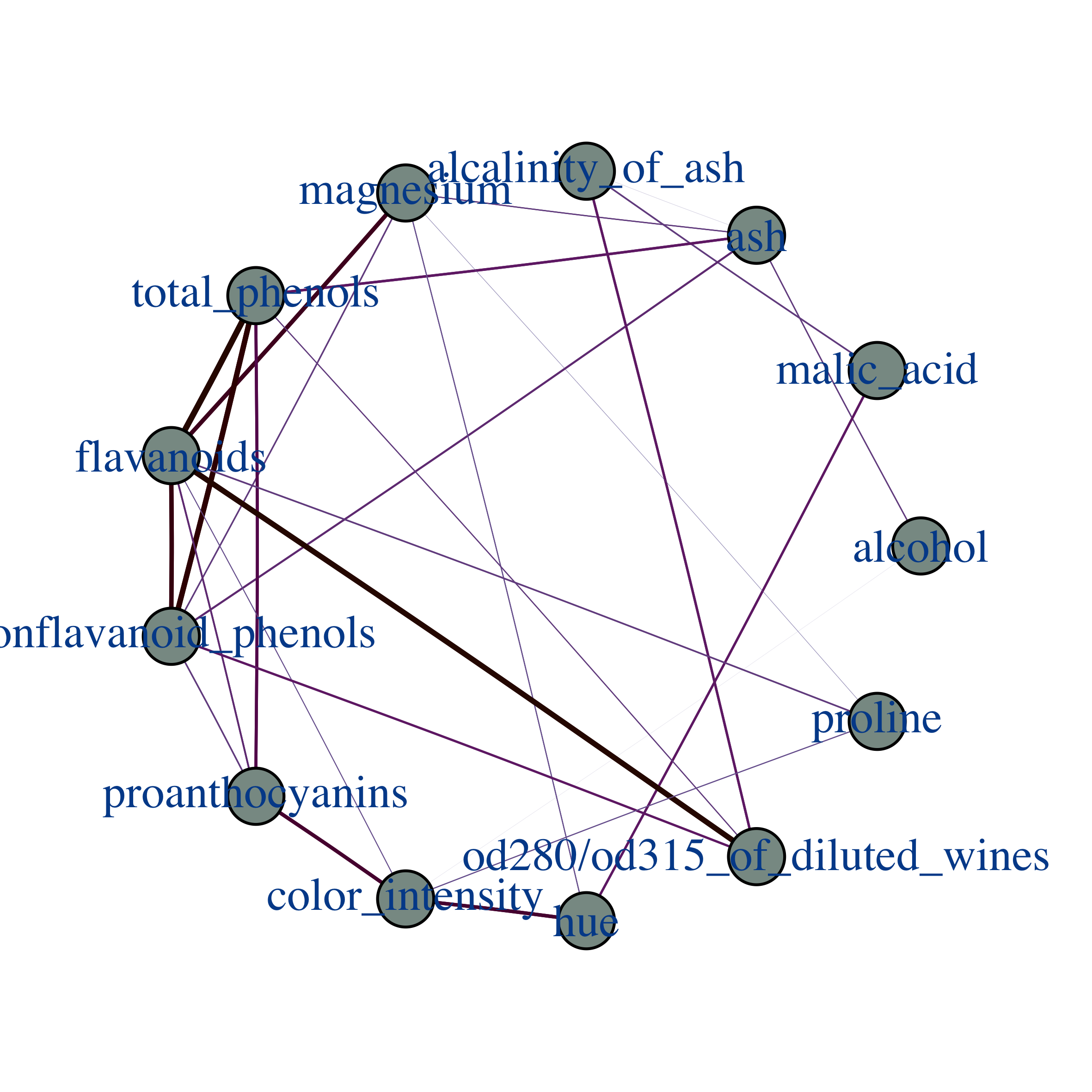}
         \caption{DN between class 2 and 3 wine networks.}
         \label{fig:bayes_dn_wine_class23}
    \end{subfigure}&    
\begin{subfigure}[c]{0.28\textwidth}
      \includegraphics[width=\textwidth]{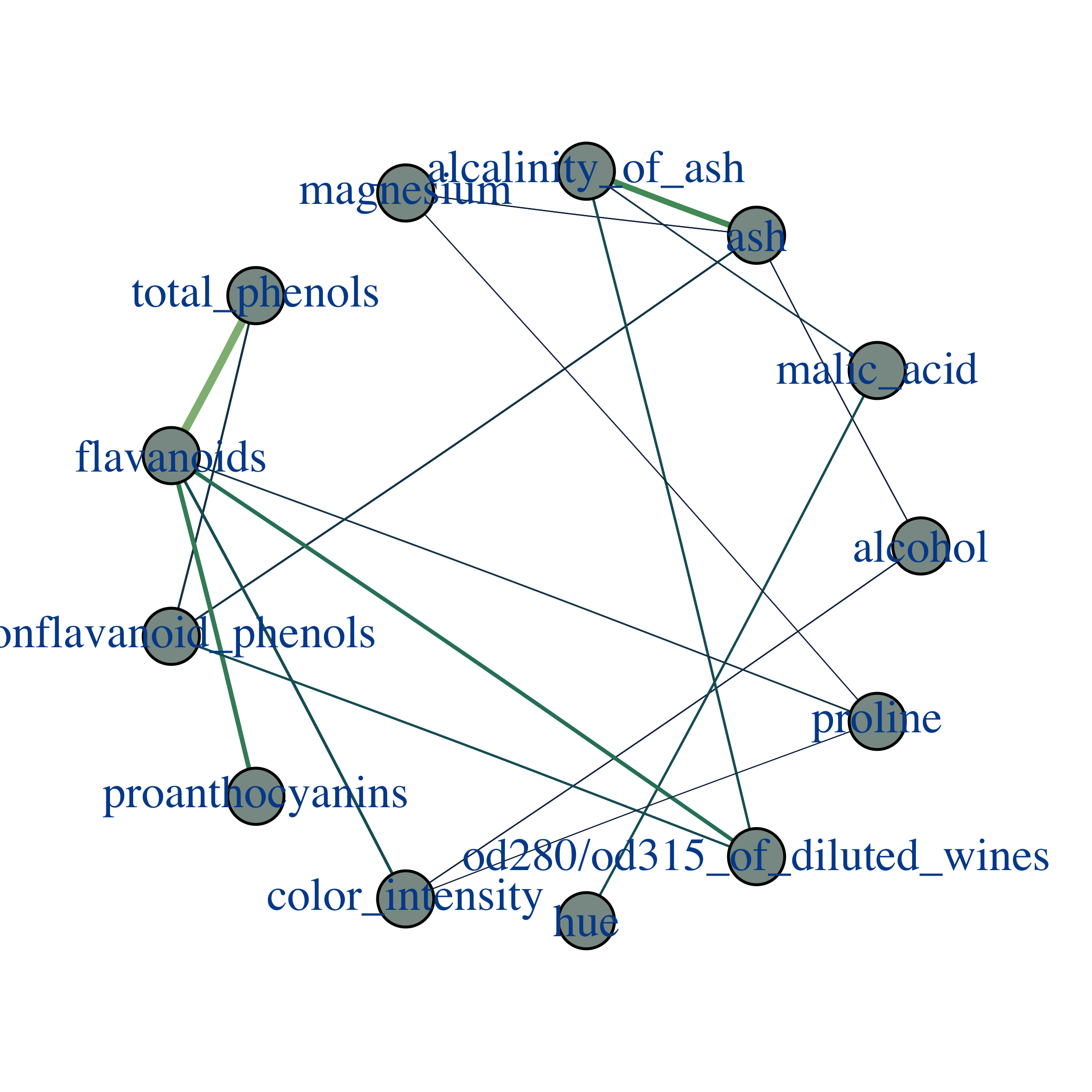}
         \caption{Class 2 wine precision matrix network.}
    \end{subfigure}&    
\begin{subfigure}[c]{0.28\textwidth}
      \includegraphics[width=\textwidth]{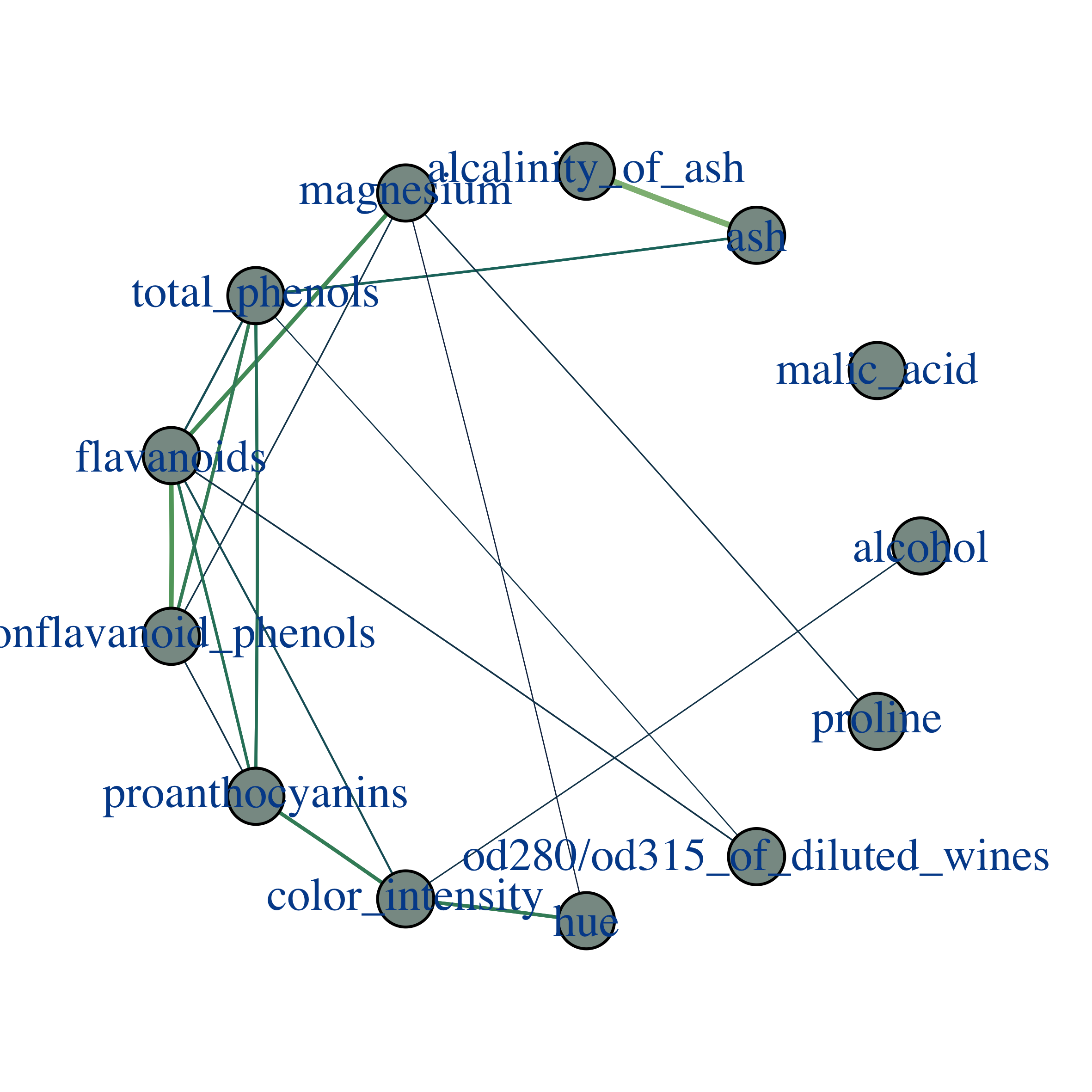}
         \caption{Class 3 wine precision matrix network.}
    \end{subfigure}         
\end{tabular}
\vspace{5mm}
\caption{
This figure presents the Bayesian DN and the corresponding precision matrix networks for each pair of wine classes: class 1 and 2, class 1 and 3, and class 2 and 3. The differential networks elucidate distinct associations between chemometric variables for each pair of wine classes.}

\label{tab:covid_networks}
\end{table}

\section{Discussion}\label{sec:discussion}
The naïve Bayesian adaptive graphical elastic net estimator exhibits excellent flexibility in the estimation of diverse topological differential networks. The estimator retains the computational efficiency and robustness of the Bayesian adaptive graphical lasso without the need for additional hyperparamters. The data driven heuristic structure learning procedure appears to perform well and unlike the approach in \citep{smith2022empowering}, does not require a priori knowledge on the anticipated network topology nor does it require sampling from a G-Wishart as in \citep{wang_2012_efficient}. The application of the BAE estimator to data encompassing breast cancer, chronic kidney disease and a classical Italian wine cultivar suggest that the estimator has the capacity and adaptability to navigate complex, interactive multidimensional applications across distinct scientific domains. Unlike frequentist based DN estimation techniques, the methodology described here offers a statistical inferential advantage through the posterior distribution results provided by the MCMC sampler. \par
The BAE estimator is not without fault and future work endeavours to address these caveats. Firstly, for posterior inference on network topology, a fully Bayesian treatment for structure learning is required. Second, a thorough investigation is required to understand the effect (if any) of "double shrinkage problem" here. An approach similar to that of Bayesian elastic net in \citep{li2010bayesian}, where their Gibbs sampler selects the $\ell_1$ and $\ell_2$ penalty parameters simultaneously. Last but not least, the blue sky goal is to develop a generalization of the BAE estimator that is capable of fully estimating the DN through simultaneous estimation of its components, $\mathbf{\Omega_1}$ and $\mathbf{\Omega_2}$, for high dimensional use cases.

\section*{Funding}
This work was based upon research supported in part by the National Research Foundation (NRF) of South Africa, NRF ref. SRUG2204203865 Ref.: RA171022270376, grant No: 119109, Ref.: RA211204653274 grant No. 151035. The opinions expressed and conclusions arrived at are those of the authors and are not necessarily to be attributed to the NRF. Mohammad Arashi's work is based upon research funded by the Iran National Science Foundation (INSF) grant No. 4015320.

\subsection{Disclosure of interest}\label{subsec:disclosure_of_interest}
The authors report no conflict of interest.

\bibliographystyle{Perfect}

\bibliography{bibliography}

\appendix{}
% \section{Analytical details}\label{app:analytical_details}
% \subsection{The Bayesian graphical elastic net block Gibbs sampler}\label{app:subsec:trunc_gaus_definition} 

% Let $X$ be a random variable following a truncated Gaussian distribution with mean $\mu$, standard deviation $\sigma$, and truncation interval $[a, b]$. The probability density function of $X$ is given by:

% \begin{equation*}
% f(x) = \frac{1}{\sigma Z} \cdot \exp\left(-\frac{(x-\mu)^2}{2\sigma^2}\right) \cdot \mathds{1}_{\{x \in [a,b]\}},
% \end{equation*}

% \noindent where $Z = \Phi\left(\frac{a - \mu}{\sigma}\right) - \Phi\left(\frac{b - \mu}{\sigma}\right)$, and $\Phi(\cdot)$ is the Gaussian cumulative distribution function. The indicator function $\mathds{1}_{\{x \in [a,b]\}}$ equals 1 if $x$ is in the interval $[a,b]$, and 0 otherwise.

\end{document}